\documentclass[
    reprint,
    amsmath,
    amssymb,
    aps,onecolumn,prx
]{revtex4-1}

\usepackage{graphicx}
\usepackage{hyperref}
\usepackage{amsmath}
\usepackage{xcolor}
\usepackage{bbm,amssymb}
\usepackage{braket}
\usepackage{bm}
\newcommand{\beq}{\begin{equation}}
\newcommand{\eeq}{\end{equation}}

\begin{document}

\title{Dynamically corrected gates from geometric space curves}

\author{Edwin Barnes}
\email{efbarnes@vt.edu}
\author{Fernando A. Calderon-Vargas}
\author{Wenzheng Dong}
\author{Bikun Li}
\author{Junkai Zeng}
\author{Fei Zhuang}

\address{Department of Physics, Virginia Tech, Blacksburg, Virginia 24061, USA}

\begin{abstract}
Quantum information technologies demand highly accurate control over quantum systems. Achieving this requires control techniques that perform well despite the presence of decohering noise and other adverse effects. Here, we review a general technique for designing control fields that dynamically correct errors while performing operations using a close relationship between quantum evolution and geometric space curves. This approach provides access to the global solution space of control fields that accomplish a given task, facilitating the design of experimentally feasible gate operations for a wide variety of applications.
\end{abstract}

\maketitle
%
%
%
%
%

\section{Introduction}

Leveraging the power of quantum mechanics to realize novel technologies capable of performing tasks far beyond present-day means is the central goal in the fields of quantum computing, sensing, and communication~\cite{GisinRMP2002,Gisin_NatPhoton2007,Ladd_Nature2010,DevoretScience2013,DegenRMP2017}. These goals demand the ability to control and entangle microscopic quantum systems with unprecedented accuracy, a task that is particularly challenging due to unwanted interactions with the surrounding environment~\cite{Chirolli_AIP2008}. Such interactions cannot be removed completely through careful system engineering since some contact with the environment is necessary in order to manipulate the quantum system. Therefore, achieving the requisite level of control requires the development of control protocols capable of coherently manipulating the systems while simultaneously mitigating the deleterious effects of the environment dynamically.

The fact that it is possible to drive a system with external control pulses that are engineered to produce an automatic self-cancellation of errors due to the environment or driving imperfections, without the need for a precise knowledge of these errors, was discovered several decades ago. This concept originated in the early literature on nuclear magnetic resonance (NMR)~\cite{Hahn_PR50,Carr_Purcell,Meiboom_Gill,Haeberlen,Vandersypen_RMP05}, where square or delta-function pulse sequences such as Hahn spin echo and the Carr-Purcell-Meiboom-Gill (CPMG) sequence quickly became indispensable tools for extending the coherence of nuclear spins for a variety of applications such as magnetic resonance imaging. These techniques have been extended to more recent contexts such as quantum information processing, where NMR sequences such as Hahn echo and CPMG have proven effective at preserving the information stored in qubits~\cite{Bluhm_NP11,Tyryshkin_NatMat11,Poletto_PRL12,Muhonen_NatNano14,Malinowski_NatNano17}. These newer applications have also driven the search for additional sequences that can more efficiently extend qubit lifetimes~\cite{Viola_PRA98,Khodjasteh_PRL05,Uhrig_PRL07,Zhang_Viola_PRB08,Jones_NJP12}.

There has also been substantial progress in developing control schemes that not only remove errors but also simultaneously rotate the quantum state of the system in some desired way~\cite{Levitt_1986,Goelman_JMR89,Wimperis1994,Cummins_PRA03,Biercuk_Nature09,Khodjasteh_PRL10,Jones2010,vanderSar_Nature12,Wang_NatComm12,Green_NJP13,Kestner_PRL13,Calderon-Vargas2016}. The analytical  tractability of ideal pulse waveforms such as delta-functions and square pulses make them attractive as building blocks in such methods. However, the use of such waveforms can also potentially limit their applicability. This is because idealized waveforms are experimentally infeasible in many quantum systems, where the need for ultrafast microsecond or nanosecond control pushes the limits of state-of-the-art waveform generators to the point where these pulse shapes cannot be reliably created, incurring large driving errors. Moreover, restricting to the use of only a few specialized pulse shapes can lead to unnecessarily long pulse sequences that quickly run up against limitations set by additional decoherence or loss mechanisms.

A further challenge that often arises when designing gate operations is the existence of physical restrictions on the range of control field amplitudes. As an example, consider the case of entangling gates between spin qubits realized through an exchange interaction (as is commonly used in electron spin-based quantum computing~\cite{Petta_Science05}). The fact that such exchange interactions are typically non-negative often rules out protocols that assume the control field can be tuned to both positive and negative values. Most NMR techniques in fact make this assumption because they are based on time-dependent magnetic field control, where such a requirement is easily met. This problem was solved in a series of publications~\cite{Wang_NatComm12,Kestner_PRL13,Wang_PRB14} that introduced a new method known as {\sc supcode}. It was shown that it is possible to generate arbitrary spin operations while removing the leading-order errors due to both charge noise and nuclear spin noise. This is achieved using specially designed sequences of square pulses, and it was further shown~\cite{Wang_PRB14} that this approach still works if the square pulses are deformed into trapezoidal shapes to allow for finite rise time restrictions. The first experimental demonstration of the {\sc supcode} sequences was performed in the context of NV centers in diamond~\cite{Rong.14}.

An additional limitation of many existing schemes for both dynamical decoupling and dynamically corrected gates is that they are usually designed around the assumption that errors are essentially static during a gate operation. This is often a reasonable starting point since error fluctuations, due to the environment or from waveform generators, are frequently found to vary slowly in time compared to the time scale of the system dynamics~\cite{Dial_PRL13,OMalley_PRApplied15,Martins_PRL16,Hutchings_PRApplied17}. However, for quantum information applications such as quantum computing which demand an unprecedented level of control accuracy, the fact that error fluctuations are not constant in time must ultimately be taken into account. Much has been learned about the structure of environmental noise in a variety of qubit systems over the past decade~\cite{Clerk2010,Dial_PRL13,Paladino2014,Martins_PRL16,Schreier_PRB08,Reilly_PRL08,Medford_PRL12,Sank_PRL12,Anton_PRB12,Sigillito_PRL15,Kalra_RSI16}, and this information can be used to further refine error-suppressing control schemes.

\begin{figure}
\centering
\includegraphics[width=0.7\columnwidth]{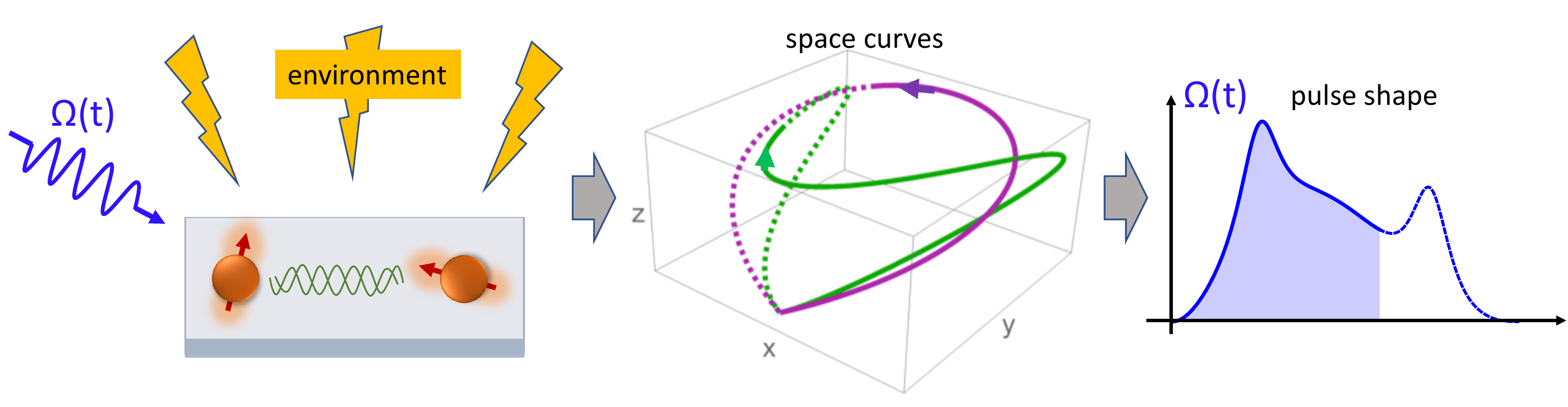}
\caption{Geometric space curves provide a general method to design quantum gates that are robust to environmental noise and other sources of error. The evolution of a system subject to noise can be mapped onto a space curve. We can reverse engineer this evolution by starting from a space curve and extracting the control Hamiltonian from its generalized curvatures. Choosing the space curve to be closed yields noise-cancelling control pulses.} 
\label{fig:cartoon}
\end{figure}

This Review describes a recently developed framework for designing dynamically corrected gates that we call Space Curve Quantum Control (SCQC). This framework relies on a geometric structure underlying the Schrödinger equation that can be exploited to overcome limitations of existing approaches. In the SCQC method, one visualizes the evolution error caused by noise as a geometric space curve. This curve lives in a space of operators that depends on the form of the control Hamiltonian and on the way in which the noise affects the system. As the system evolves in time, the curve winds through this space with constant velocity. The net displacement between the initial and final points of the curve quantify the deviation from the system's ideal evolution. Any dynamically corrected gate therefore corresponds to a closed space curve, providing a global view of the solution space of robust gates. We show how one can systematically design robust gate operations by starting from closed space curves and computing control fields from their generalized curvatures. The general strategy is illustrated in Fig.~\ref{fig:cartoon}. As we describe in this work, this approach can be applied to a variety of contexts, including the design of single- or multi-qubit gates and Landau-Zener interferometry, both for quasistatic and time-dependent noise. Moreover, it can be combined with holonomic methods to suppress multiple noise sources simultaneously. Space curves also provide a natural way to obtain dynamically corrected gates that operate near the quantum speed limit. While here we focus on correcting noise errors, the method can be applied to any situation in which reverse-engineering the evolution of a quantum system is needed.

This Review is organized as follows. In Sec.~\ref{sec:SCQC1qubit}, we show how the evolution of a driven qubit subject to noise maps to closed space curves in two or three dimensions. We also show how to obtain time-optimal dynamically corrected gates by finding curves of minimal length. In Sec.~\ref{sec:LandauZener}, we adapt the SCQC formalism to the Landau-Zener problem in which the energy gap at the avoided crossing fluctuates. We show that non-monotonic sweeps through the avoided crossing can suppress noise errors while performing operations, while monotonic sweeps cannot. We also present a general recipe for constructing closed curves of constant torsion, which are the curves that describe Landau-Zener physics. We extend the SCQC framework to multi-level and multi-qubit systems in Sec.~\ref{sec:generalized_geometric}. There we present a general method for relating control Hamiltonians to generalized curvatures, and we also give examples of dynamically corrected gates in coupled two-qubit systems. In Sec.~\ref{sec:time-dependent}, we show that pulses which cancel low-frequency time-dependent noise can be obtained from sequences of closed curves, and we describe a systematic numerical technique for obtaining such sequences. We demonstrate that the resulting pulses are effective in suppressing $1/f$ noise, a type of noise that is ubiquitous in solid-state qubit platforms~\cite{Paladino2014}. Sec.~\ref{sec:multiple_noises} discusses the case of two noise sources afflicting a qubit. We survey several approaches to designing gates that cancel both types of noise simultaneously, focusing primarily on the recently developed ``doubly geometric" approach, which combines holonomic evolution with the SCQC framework. A discussion of the relationship between SCQC and numerical optimal control methods is given in Sec.~\ref{sec:outlook}, along with some concluding remarks.

\section{Dynamically corrected gates from space curves}\label{sec:SCQC1qubit}

\subsection{Resonantly driven qubit and plane curves}\label{sec:2d_curves}

We begin by illustrating how the SCQC formalism works in the simplest example: a qubit driven by a single control field and subject to stochastic noise that is transverse to the driving field axis. The Hamiltonian is
\beq\label{eq:Hamresdriving}
\mathcal{H}=\frac{\Omega(t)}{2}\sigma_x +\epsilon\sigma_z,
\eeq
where $\sigma_x$ and $\sigma_z$ are Pauli matrices, and $\epsilon$ represents a stochastic fluctuation in the energy levels of the qubit, and $\Omega(t)$ is the envelope of the driving field. In the context of a qubit driven by a laser or by ac electric or magnetic fields, this Hamiltonian corresponds to resonant driving in the absence of the noise error $\epsilon$. This model was used to derive many of the classic dynamical decoupling protocols, including Hahn spin echo, CPMG, etc. This can be done by expanding the evolution operator $\mathcal{U}(t)$ that is generated by $\mathcal{H}$ in powers of $\epsilon$:
\beq
\mathcal{U}(t)=\sum_n \epsilon^n U_n(t).
\eeq
One finds that $U_n(t)$ for $n>0$ is a matrix that depends solely on the complex function
\beq
g_n(t)=\int_0^tdt'e^{i\int_0^{t'}dt''\Omega(t'')}g_{n-1}^*(t').\label{eq:curve_constraints}
\eeq
Note that this function is defined recursively order by order, with $g_0(t)=1$. Removing the error at order $n$ is tantamount to requiring $U_n(T)=0$, where $T$ is the final time/gate duration. This in turn requires $g_n(T)=0$. Our goal therefore is to find choices of the pulse profile $\Omega(t)$ that satisfy this condition. Dynamical decoupling sequences like spin echo or CPMG can be derived by choosing an ansatz for $\Omega(t)$ comprised of a superposition of delta-function pulses and then solving $g_n(T)=0$ to determine the times at which the pulses should be applied. 

\begin{figure}
\centering
\includegraphics[width=0.5\columnwidth]{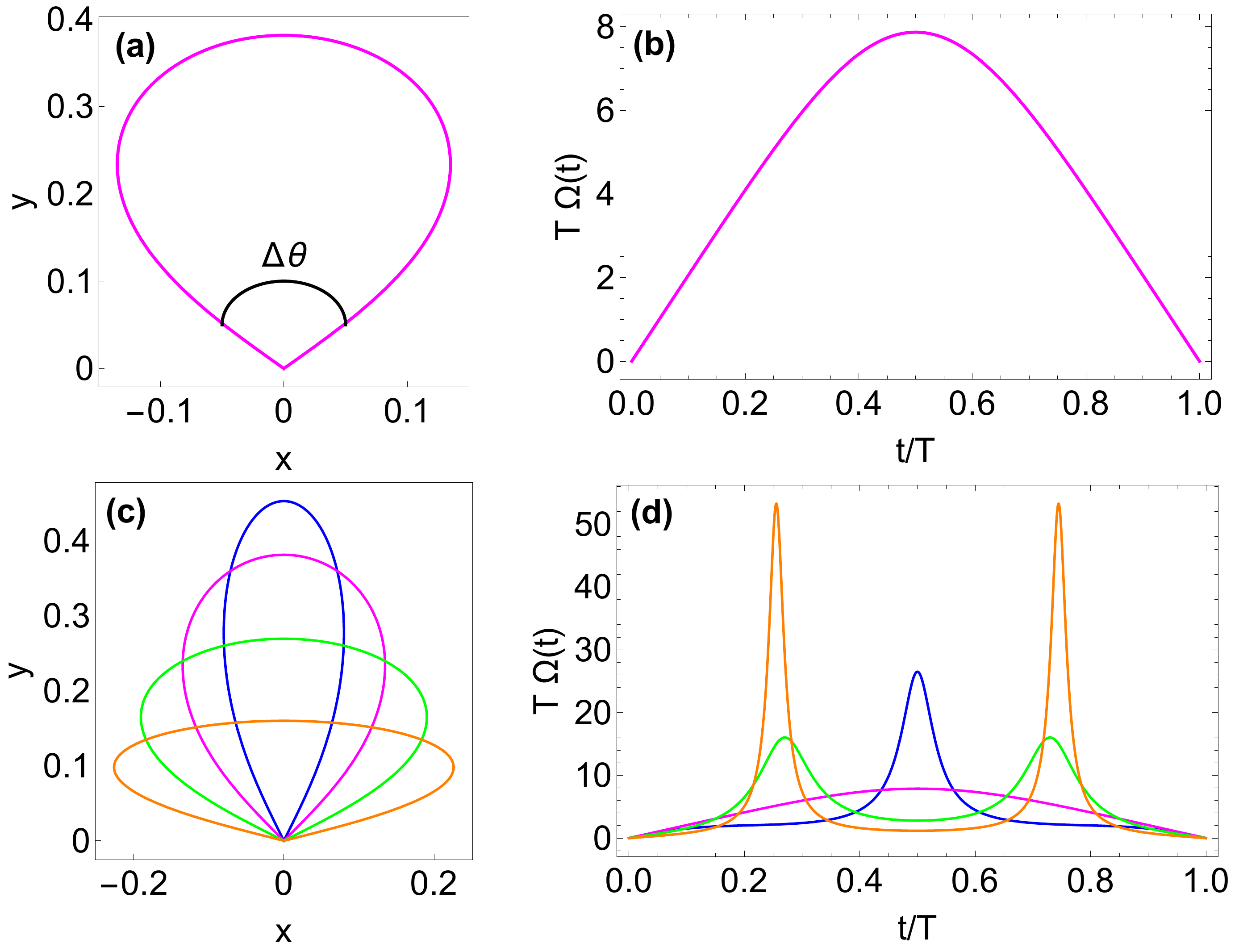}
\caption{(a) Example of a graphical solution to the first-order transverse noise constraint $g_1(T)=0$, and (b) the corresponding driving field that implements error-suppressed evolution. The opening angle $\Delta\theta$ of the curve in (a) determines the angle of rotation $\varphi$ of the gate operation, the length of the curve is the duration of the pulse, and its extrinsic curvature gives the driving field shown in (b). (c) Several curves satisfying the first-order error cancellation condition with different values of $\Delta\theta$ and (d) their corresponding pulses. This figure is adapted from~\cite{Zeng_NJP2018}.}
\label{fig:junkai1}
\end{figure}

\begin{figure}
\centering
\includegraphics[width=0.5\columnwidth]{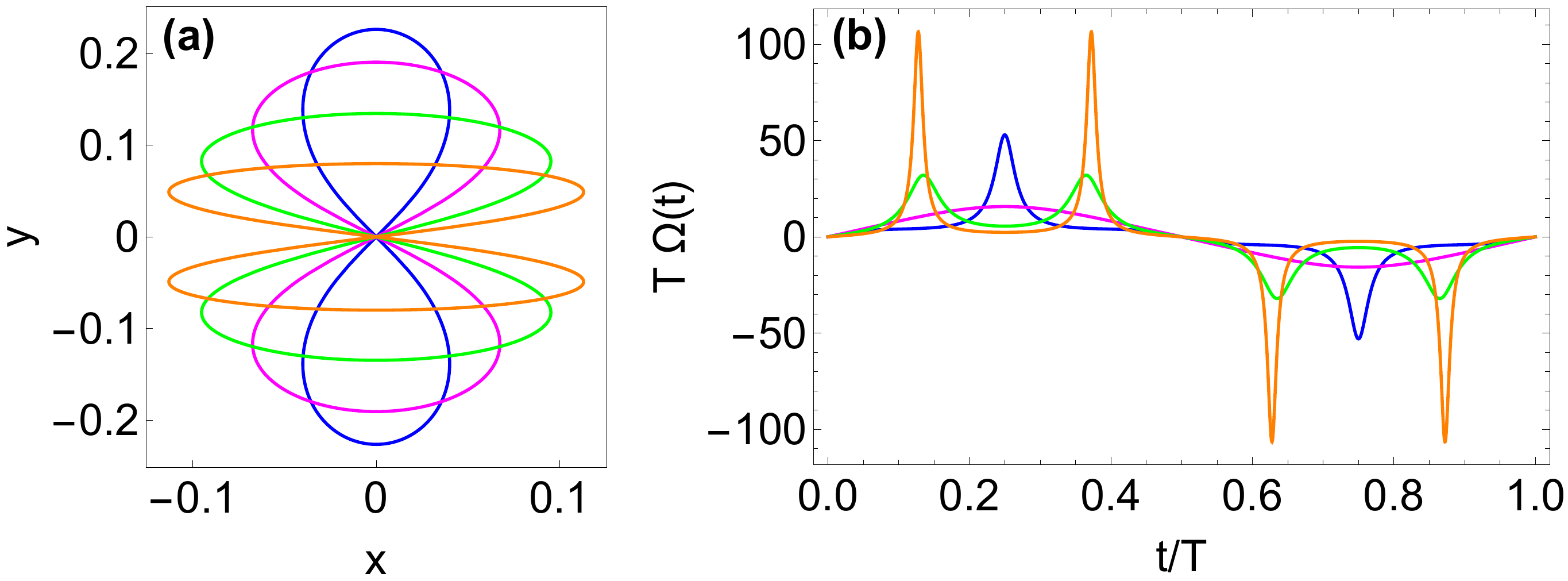}
\caption{(a) Examples of graphical solutions to the second-order transverse noise constraint $g_2(T)=0$, and (b) the corresponding driving fields that implement error-suppressed evolution. Any closed curve with zero net area gives a pulse that cancels errors to second order. This figure is adapted from~\cite{Zeng_NJP2018}.}
\label{fig:junkai2}
\end{figure}

Ref.~\cite{Zeng_NJP2018} showed that the constraints $g_n(T)=0$ admit simple geometrical interpretations that reveal the most general solution to this problem. The starting point is to notice that since $g_1(t)$ is a complex function, it can be thought of as describing a curve in a two-dimensional plane spanned by $\mathrm{Re}(g_1)$ and $\mathrm{Im}(g_1)$. The curve starts at the origin at $t=0$ (since $g_1(0)=0$) and traces out a path as time evolves. In this picture, we can interpret the constraint $g_1(T)=0$ as the condition that this curve comes back to the origin and closes on itself at time $T$. Thus, driving fields which cancel the first-order constraint are in one-to-one correspondence with closed curves in the plane. Furthermore, it was shown that the opening angle of the curve at the origin determines the angle of the rotation that is implemented by $\Omega(t)$. Examples of such curves are shown in Fig.~\ref{fig:junkai1}(a,c). Additional examples were also presented in Ref.~\cite{deng2021correcting}.

\begin{figure}
    \centering
    \includegraphics[width=\columnwidth]{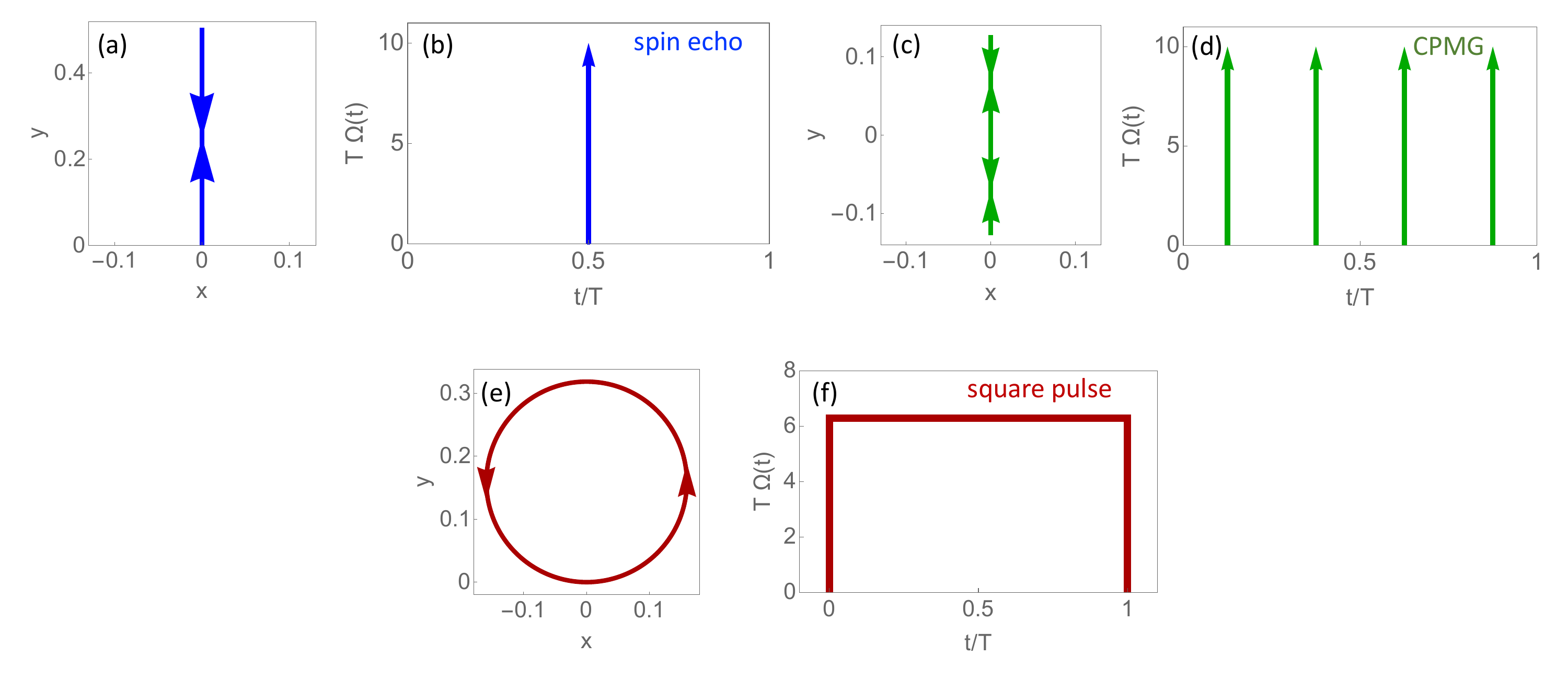}
    \caption{Delta function and square pulses in the SCQC formalism. (a) A straight line that starts and ends at the origin, retracing itself in the second half of the evolution corresponds to (b) a spin echo $\pi$ pulse. (c) A straight line that starts and ends at the origin, retracing itself multiple times yields (d) the CPMG sequence. (e) A circle corresponds to (f) a square pulse. This figure is adapted from \cite{Zeng_NJP2018}.}
    \label{fig:ideal_pulses}
\end{figure}

Driving fields which also satisfy the second-order constraint $g_2(T)=0$ again correspond to curves in a two-dimensional plane that start and end at the origin, but now they must enclose a region that has zero net area. This follows from the observation that $g_2(T)$ is proportional to the enclosed area. Examples of such curves are shown in Fig.~\ref{fig:junkai2}(a). These planar curves have a built-in orientation, and for self-crossing curves as in Fig.~\ref{fig:junkai2}(a), this orientation is opposite in the two loops. Thus, the areas enclosed by the two loops exactly cancel. Geometrical interpretations of the third- and fourth-order constraints in terms of signed volumes over the plane curve were also discovered in~\cite{Zeng_NJP2018}.

A remarkable feature of this geometrical framework is that {\it the pulse shape is the curvature of the curve}: 
\beq
\Omega(t)=\dot x\ddot y-\ddot x\dot y.
\eeq
Here, we have expressed the curve as $\bm{r}(t)=x(t)\hat x+y(t)\hat y$, where $x=\mathrm{Re}(g_1)$ and $y=\mathrm{Im}(g_1)$. This formula can be confirmed easily by computing the derivatives of $g_1(t)$ using its definition, Eq.~(\ref{eq:curve_constraints}). Furthermore, {\it time is the arc-length parameterization of the curve}: $\|\dot{\bm{r}}(t)\|=1$. This is a direct consequence of the fact that $\dot g_1(t)$ is a pure phase.
This simple relationship between plane curves and robust pulse shapes facilitates the process of producing experimentally feasible control waveforms. Moreover, this approach provides a global view of the solution space since any noise-cancelling pulse can be obtained from the curvature of a closed curve. 

Because this framework is general, it must also include all previously known dynamical decoupling and dynamically corrected gate sequences developed for quasi-static noise. This is indeed the case. As illustrated in Fig.~\ref{fig:ideal_pulses}, delta-function sequences translate to curves that lie on a line. In the case of Hahn spin echo~\cite{Hahn_PR50}, where a single delta-function $\pi$ pulse is applied halfway through the evolution, the curve starts at the origin and extends outward linearly since the curvature is zero. The curve turns around at the midpoint of the evolution ($t=T/2$) and then retraces its path, returning to the origin at time $t=T$ (Fig.~\ref{fig:ideal_pulses}(a)). At the midpoint, the curvature is infinite, which corresponds to the delta-function pulse (Fig.~\ref{fig:ideal_pulses}(b)). Any other sequence of delta-function $\pi$ pulses that cancels noise also maps to a straight line, although now the line is retraced multiple times, and a $\pi$ pulse is included in the sequence each time the curve turns around. This is illustrated for a 4-pulse CPMG sequence in Fig.~\ref{fig:ideal_pulses}(c,d). On the other hand, sequences based on square pulses correspond to curves comprised of connected circular arcs, since a constant pulse maps to a curve of constant curvature. A complete circle corresponds to a square pulse that implements an identity operation while cancelling noise to first order (see Fig.~\ref{fig:ideal_pulses}(e,f)). In Ref.~\cite{Zeng_NJP2018}, it was pointed out that, for fixed $T$, the pulses must become more sharply peaked as the order of noise cancellation is increased. This finding is consistent with a theorem proven in Ref.~\cite{Wang_NatComm12} which states that noise cancellation at arbitrarily high orders cannot be achieved with smooth pulses. In practical implementations, however, it is typically not necessary to go beyond the first few orders in order to achieve sufficient control accuracy.

\subsection{Time-optimal pulses from short plane curves}

The fact that in the SCQC framework the length of the curve is equal to the evolution time provides a powerful mechanism to find the fastest possible pulses that implement operations while cancelling noise, as was first shown in Ref.~\cite{Zeng_PRA18}. Experimental constraints on the pulse shape must be taken into account, because otherwise the fastest pulses will be delta functions. This is because the gate time can always be reduced by shrinking the curve while keeping its shape fixed, at the expense of increasing the curvature, and hence the pulse amplitude. Of course, in any real physical qubit system, there will be a limit on the driving power that can be applied, and this is taken into account by placing a restriction on the pulse amplitude:
\begin{equation}
	|\Omega(t)|\le\Omega_{\mathrm{max}},\label{eq:ampconstraint}
\end{equation}
for some constant $\Omega_{\mathrm{max}}$.

\begin{figure}
 \centering
\includegraphics[width=0.7\columnwidth]{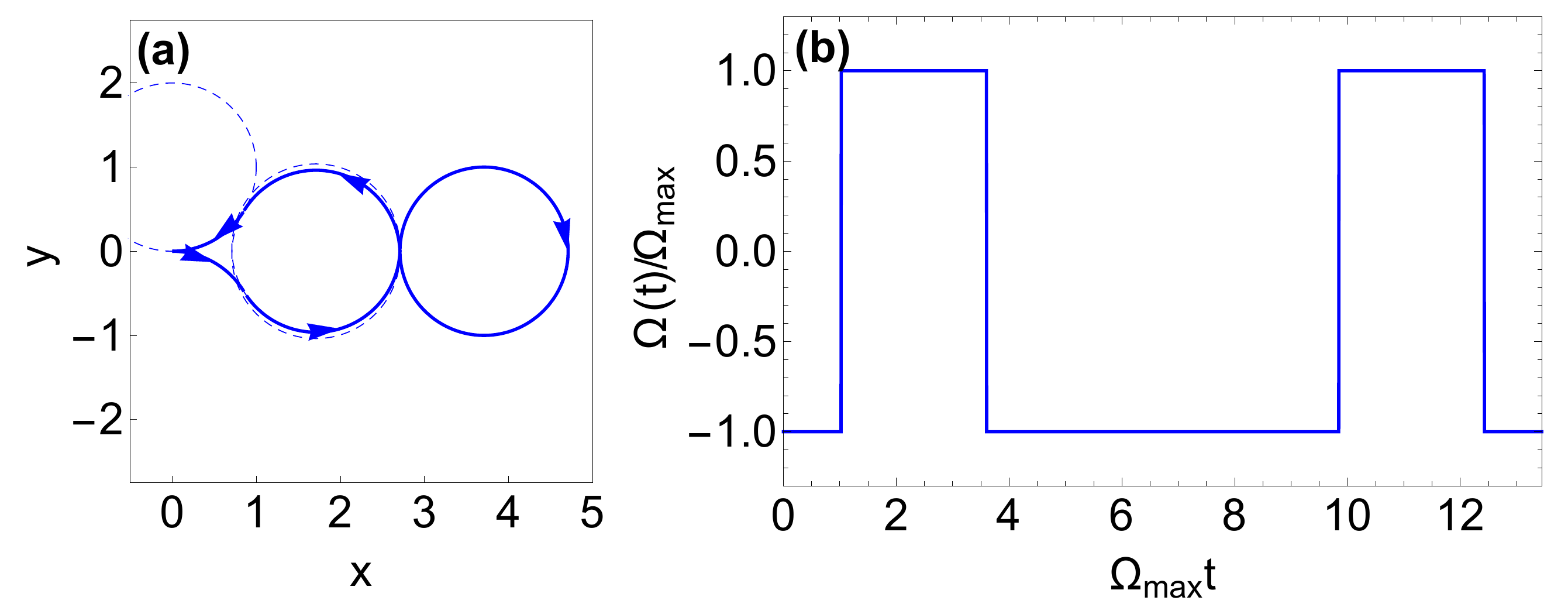}
\caption{(a) Curve that yields the fastest single-qubit $\pi$-rotation while canceling noise to second order and respecting the pulse constraint $|\Omega(t)|\le\Omega_{\mathrm{max}}$. (b) Corresponding driving pulse. This figure is adapted from~\cite{Zeng_PRA18}.}
\label{fig:secondorder}
\end{figure}

Geometrically, Eq.~(\ref{eq:ampconstraint}) imposes an upper bound on the curvature of the plane curve. The problem of looking for the fastest pulses then amounts to searching for the shortest curves that respect this curvature bound while also satisfying error-cancellation constraints (closed curve, zero net area) and the target rotation constraint (angle subtended at origin). Ref.~\cite{Zeng_PRA18} solved this problem by recasting it as a variational calculus problem in which the objective function to be minimized is the curve length, with the pulse amplitude and zero-area constraints included with slack variables and Lagrange multipliers. The closed-curve condition is imposed through boundary conditions. In Ref.~\cite{Zeng_PRA18}, it was shown that there is a unique optimal solution to this problem corresponding to a curve comprised of five circular arcs connected together. An example curve that yields a $\pi$ rotation and its corresponding pulse are shown in Fig.~\ref{fig:secondorder}(a,b). Since circular arcs have constant curvature, they correspond to driving pulses of constant amplitude, i.e., square pulses. Thus, the global time-optimal pulse is a composite square pulse. Similar findings were also recently obtained using the Pontryagin Maximum Principle and shortcuts to adiabaticity~\cite{Stefanatos_PRA2019,Ansel_2021}. We see that the SCQC formalism naturally provides a geometric understanding of the quantum speed limit \cite{Mandelstam_JPhys45,Margolus_PD98,Deffner_PRL13,Deffner_2017} for a given target operation in the presence of control field constraints.

Although these composite square pulses respect the experimental constraint of finite pulse amplitude, they are not yet experimentally practical because they require infinitely fast pulse rise times. However, they still constitute a good starting point for designing smooth pulses that accomplish the same tasks at speeds close to the global optimum. In Ref.~\cite{Zeng_PRA18}, two approaches to producing smooth pulses were presented: one based on smoothing the square waveform, and one based on first applying a smoothing procedure to the curve itself before extracting the pulse from its curvature. It was found that the latter approach provides better pulses, because it is easier to maintain the noise-cancellation conditions if one works directly with the curve.

\subsection{Arbitrary single-qubit gates from space curves in three dimensions}\label{sec:non-resonant}

All the results described so far apply to the case of single-axis (resonant) driving, Eq.~(\ref{eq:Hamresdriving}). In subsequent work~\cite{Zeng_PRA19}, it was discovered that additional geometrical structure hidden within the time-dependent Schr$\ddot{\hbox{o}}$dinger equation allows one to obtain all robust driving fields in the most general case of three-axis driving. In this case, the Hamiltonian of a driven two-level system subject to a single source of quasistatic noise can be expressed as
\begin{equation}
\mathcal{H}(t)=\mathcal{H}_c(t)+\delta\mathcal{H}\\
=\frac{\Omega(t)\cos\Phi(t)}{2}\sigma_x+\frac{\Omega(t)\sin\Phi(t)}{2}\sigma_y+\frac{\Delta(t)}{2}\sigma_z+\epsilon\sigma_z,
\label{eq:hamil}
\end{equation}
where $\Omega(t)$, $\Phi(t)$, and $\Delta(t)$ are three independent driving fields. $\delta\mathcal{H}=\epsilon\sigma_z$ is the quasistatic noise term, which is assumed to be weak compared to the maximum amplitudes of the driving fields $\Omega(t)$ and $\Delta(t)$. Ref.~\cite{Zeng_PRA19} showed that the evolution operators generated by Hamiltonians of the form of ${\cal H}_c(t)$ in (\ref{eq:hamil}) are in one-to-one correspondence with geometric space curves in three dimensions. The space curve is defined in terms of the integral of the interaction-picture Hamiltonian:
\begin{equation}\label{eq:space_curve_def}
    \int^t_0\mathcal{U}_c^{\dagger}(t')\sigma_z\mathcal{U}_c(t')dt'=\bm{r}(t)\cdot \bm{\sigma} =  x(t)\sigma_x+y(t)\sigma_y+z(t)\sigma_z,
\end{equation}
where $\mathcal{U}_c(t)$ is the evolution operator generated by ${\cal H}_c(t)$. Any space curve in three dimensions is characterized by two real functions known as the curvature and torsion. The curvature $\kappa(t)$ quantifies how quickly the tangent vector, $\dot{\bm{r}}(t)$, is changing direction at each point along the curve, while the torsion $\tau(t)$ provides a measure of how quickly the curve is bending out of a plane spanned by the tangent vector and its derivative. In the SCQC formalism, the curvature and torsion are determined by the driving fields:
\begin{eqnarray}
\label{eq:curvature_torsion}
    \kappa(t)&=& \|\ddot{\bm{r}}(t)\| =\Omega(t),\\
 \tau(t) &=&  \frac{(\dot{\bm{r}}\times \ddot{\bm{r}})\cdot\dddot{\bm{r}} }{\big\| \dot{\bm{r}}\times \ddot{\bm{r}}  \big\|^2}  = \dot{\Phi}(t)-\Delta(t).  
\end{eqnarray}
Thus, given any space curve, the driving field $\Omega(t)$ that generates this evolution can be extracted from the curvature of the curve. However, notice that only the difference $\dot{\Phi}(t)-\Delta(t)$ is fixed by the torsion, meaning that $\Phi$ and $\Delta$ are not uniquely determined by the space curve. This is related to the fact that one can always transform to a frame in which one of these fields is effectively eliminated. For example, performing a frame transformation $R(t)=e^{-i\int^t_0\frac{\Delta(t')}{2}\sigma_z dt'}$ converts the control Hamiltonian into $\widetilde{\cal{H}}_c=R^\dag{\cal{H}}R-iR^\dag\dot R=\frac{\Omega}{2}(\cos\widetilde{\Phi}\sigma_x+\cos\widetilde{\Phi}\sigma_y)$, where $\dot{\widetilde{\Phi}}=\dot\Phi-\Delta$. The space curve therefore encodes the qubit evolution up to such a frame transformation. Different evolutions that are related by a frame transformation of this form all map to the same space curve.

The space curve $\bm{r}(t)$ contains full information about the evolution operator $\mathcal{U}_c(t)$. To see how to extract this information, first write $\mathcal{U}_c(t)=R(t)\widetilde{\mathcal{U}}_c(t)$, where $\widetilde{\mathcal{U}}_c(t)$ is the evolution operator generated by $\widetilde{\mathcal{H}}_c(t)$. This avoids the frame ambiguity described above and allows $\widetilde{\mathcal{U}}_c(t)$ to be uniquely determined from the space curve. To see this explicitly, parameterize the evolution operator as
\begin{equation}\label{eq:evol_parameterized}
    \widetilde{\mathcal{U}}_c(t)=\begin{pmatrix} e^{i(\zeta+\lambda)/2}\cos(\theta/2) & -i e^{-i(\zeta-\lambda)/2}\sin(\theta/2) \cr -ie^{i(\zeta-\lambda)/2}\sin(\theta/2) & e^{-i(\zeta+\lambda)/2}\cos(\theta/2)\end{pmatrix},
\end{equation}
where $\zeta$, $\lambda$, $\theta$ are all real functions of time. The tangent curve (also known as tangent indicatrix or tantrix for short) is then given by
\begin{equation}\label{eq:tantrix}
    \dot{\bm{r}}=\mathcal{U}_c^\dag\sigma_z\mathcal{U}_c=\widetilde{\mathcal{U}}_c^\dag\sigma_z\widetilde{\mathcal{U}}_c=-\sin\theta\sin\zeta\hat x+\sin\theta\cos\zeta\hat y+\cos\theta\hat z.
\end{equation}
We see that two of the three functions ($\theta$ and $\zeta$) in the evolution operator can be obtained from $\dot{\bm{r}}$, while $\lambda$ cannot. However, using the Schr\"odinger equation for $\widetilde{\mathcal{U}}_c(t)$, we can also derive a formula for this phase in the evolution operator:
\begin{equation}\label{eq:evol_phase}
    \lambda=-\int_0^tdt'\tau(t')+\arctan\left(\frac{\ddot x\dot y-\ddot y\dot x}{\ddot z}\right).
\end{equation}
Thus, the full evolution operator in the original frame can be obtained from $\int_0^tdt'\Delta(t')$ and from data extracted from the space curve: $\dot{\bm{r}}$, $\ddot{\bm{r}}$, and the integral of the torsion.

Notice that in Eq.~(\ref{eq:space_curve_def}) we defined the space curve to be the Pauli coefficients of the first-order term in the Magnus expansion of the evolution operator in the interaction picture: $\mathcal{U}_I(t)\approx\exp[-i\epsilon\int^t_0\mathcal{U}_c^{\dagger}(t')\sigma_z\mathcal{U}_c(t')dt']=\exp[-i\epsilon\bm{r}(t)\cdot\bm{\sigma}]$. In the absence of noise ($\epsilon=0$) we would have $\mathcal{U}_I(t)=\mathbbm{1}$. We see that this result can be recovered at the final time $t=T$ to first order in $\epsilon$ by requiring the space curve to be closed: $\bm{r}(T)=\bm{r}(0)=0$. Therefore, control fields that dynamically suppress noise can be obtained by drawing closed curves and extracting the curvature and torsion. The desired target evolution can be chosen by fixing the tangent vector at the end of the curve and by choosing the total torsion (i.e., the integral of the torsion) appropriately. This simple relationship between space curves and robust control fields makes the process of producing experimentally feasible pulse waveforms transparent. It is possible to obtain smooth, dynamically correcting pulses for any desired single-qubit gate with this approach. It is also important to point out that this constitutes a general solution to the problem: Any pulse that cancels noise corresponds to a closed space curve.If we restrict to Hamiltonians for which $\dot\Phi=\Delta$, then the corresponding space curves have zero torsion. In this case, the curves lie in a plane, and we recover the SCQC formalism for single-axis driving discussed in Sec.~\ref{sec:2d_curves}.

\begin{figure}
 \centering
\includegraphics[width=0.34\columnwidth]{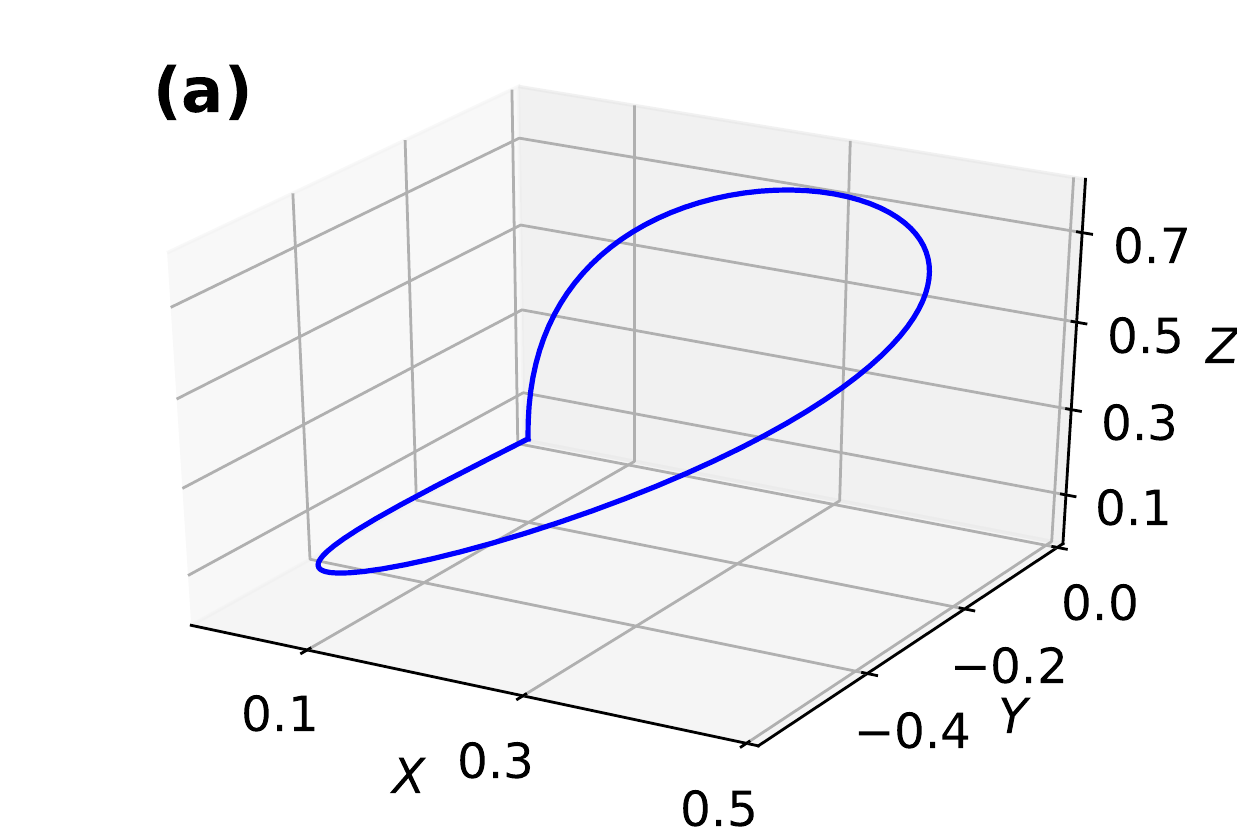}
\includegraphics[width=0.32\columnwidth]{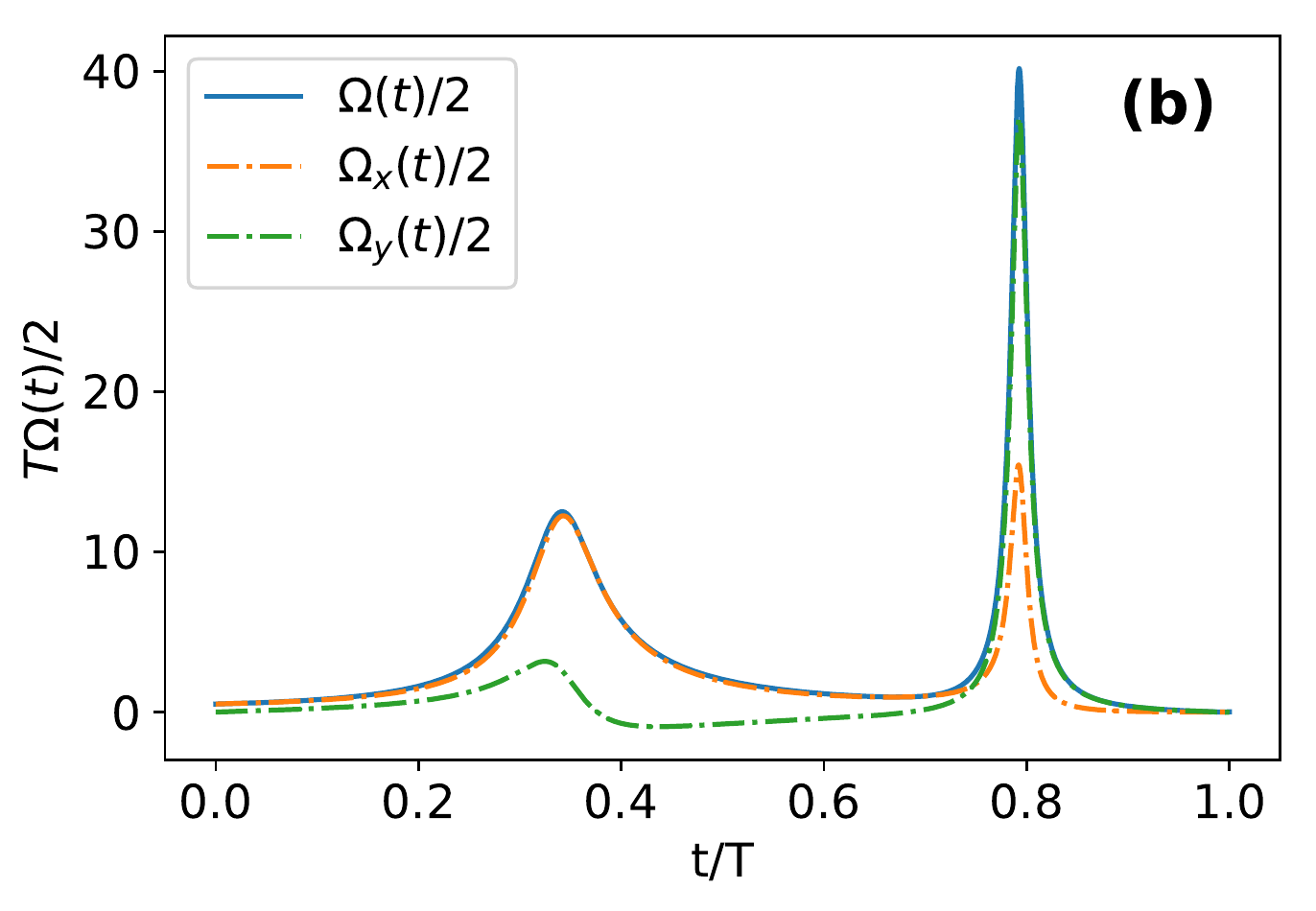}
\includegraphics[width=0.32\columnwidth]{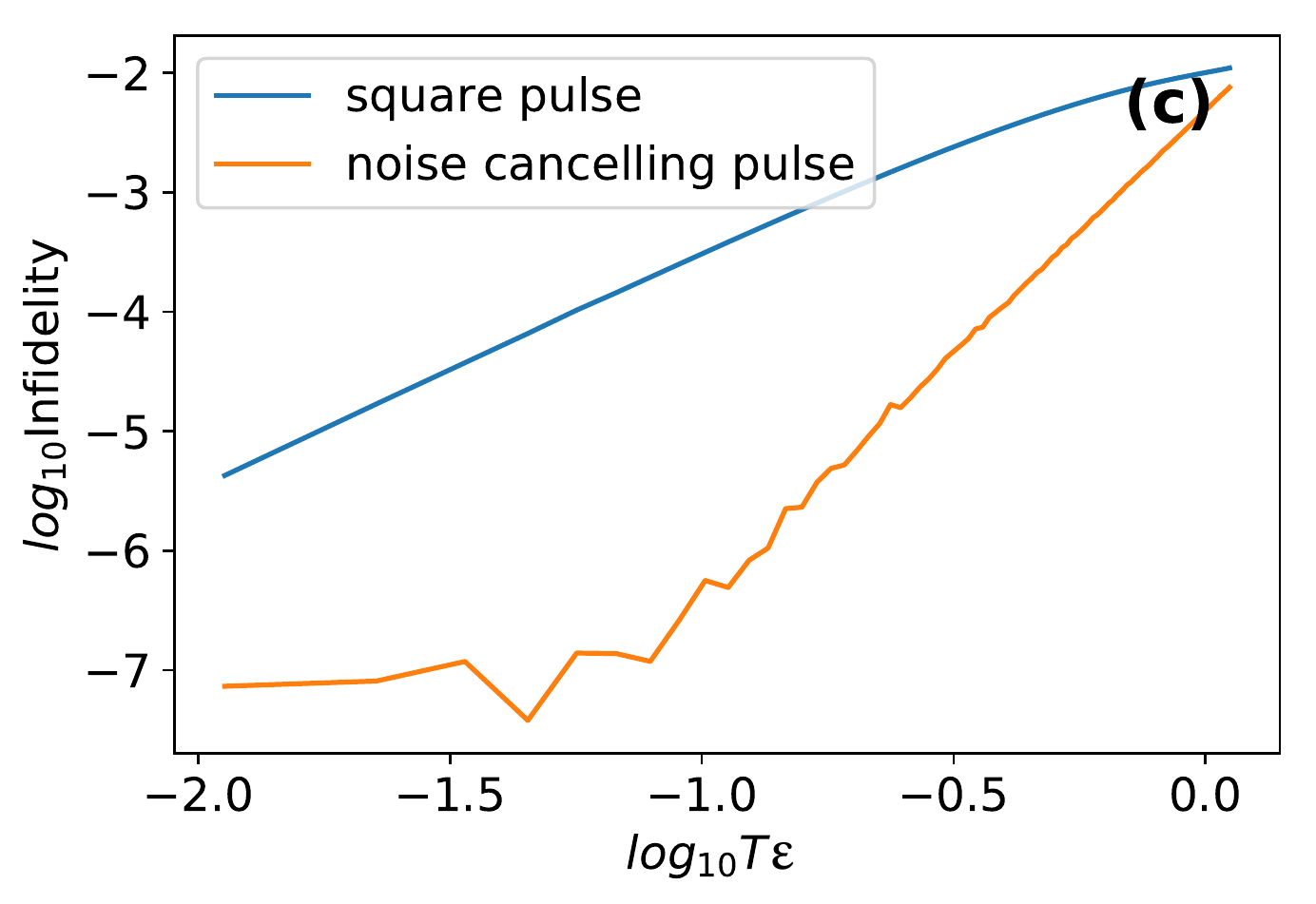}
\caption{Dynamically corrected Clifford gate $R(-\hat x+\hat y+\hat z,2\pi/3)$. (a) Closed space curve. (b) The control fields obtained from the curvature and torsion of the curve shown in (a). Here, $\Omega_x=\Omega\cos\Phi$, $\Omega_y=\Omega\sin\Phi$. (c) Comparison of the log-log infidelity for the geometrically engineered gate (orange) and for the gate implemented by a naive square pulse (blue). This figure is adapted from Ref.~\cite{Zeng_PRA19}. Further details about this example can be found in that reference.}
\label{fig:curve1}
\end{figure}
An example of how this geometrical structure can be exploited to design dynamically corrected gates is shown in Fig.~\ref{fig:curve1}. Here, the target gate operation is one of the Clifford gates: $\mathcal{U}_c(T)=R(-\hat x+\hat y+\hat z,2\pi/3)$, i.e., a rotation about the axis $-\hat x+\hat y+\hat z$ by angle $2\pi/3$. A pulse that generates this gate while canceling first-order errors can be obtained from a closed curve that has the appropriate slope as it returns to the origin, as shown in Fig.~\ref{fig:curve1}(a). The control fields extracted from the curvature and torsion are shown in Fig.~\ref{fig:curve1}(b). A plot of the infidelity of the resulting gate as a function of the noise strength is shown in Fig.~\ref{fig:curve1}(c), where for comparison, the result for a square pulse of the same duration is also shown. It is evident that the noise-suppressing pulse makes the operation orders of magnitude more robust than a naive square pulse, and the slope of the log-log infidelity plot confirms that the first-order error is cancelled.

\begin{figure}
\centering
\includegraphics[width=0.4\columnwidth]{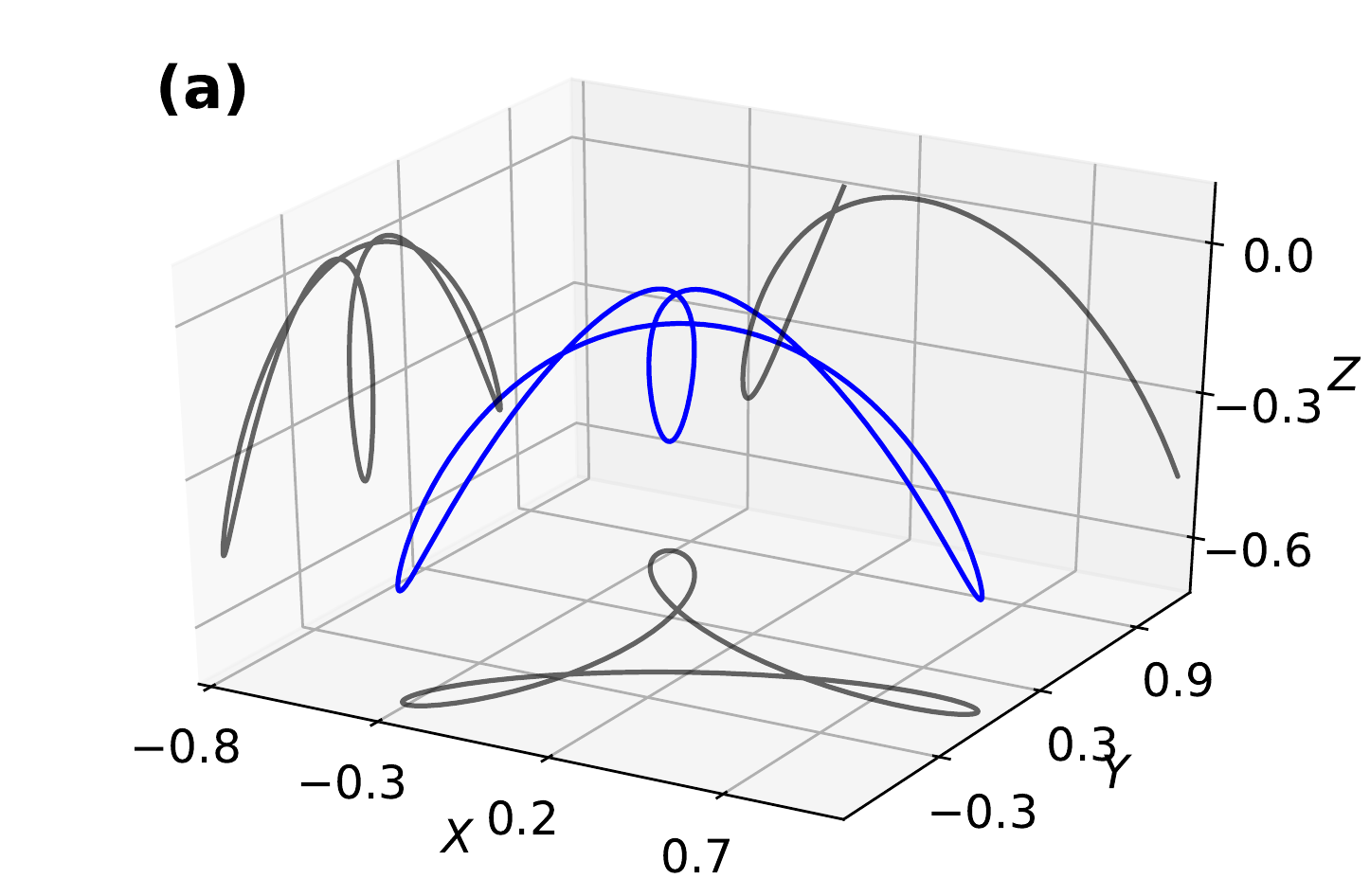}
\includegraphics[width=0.4\columnwidth]{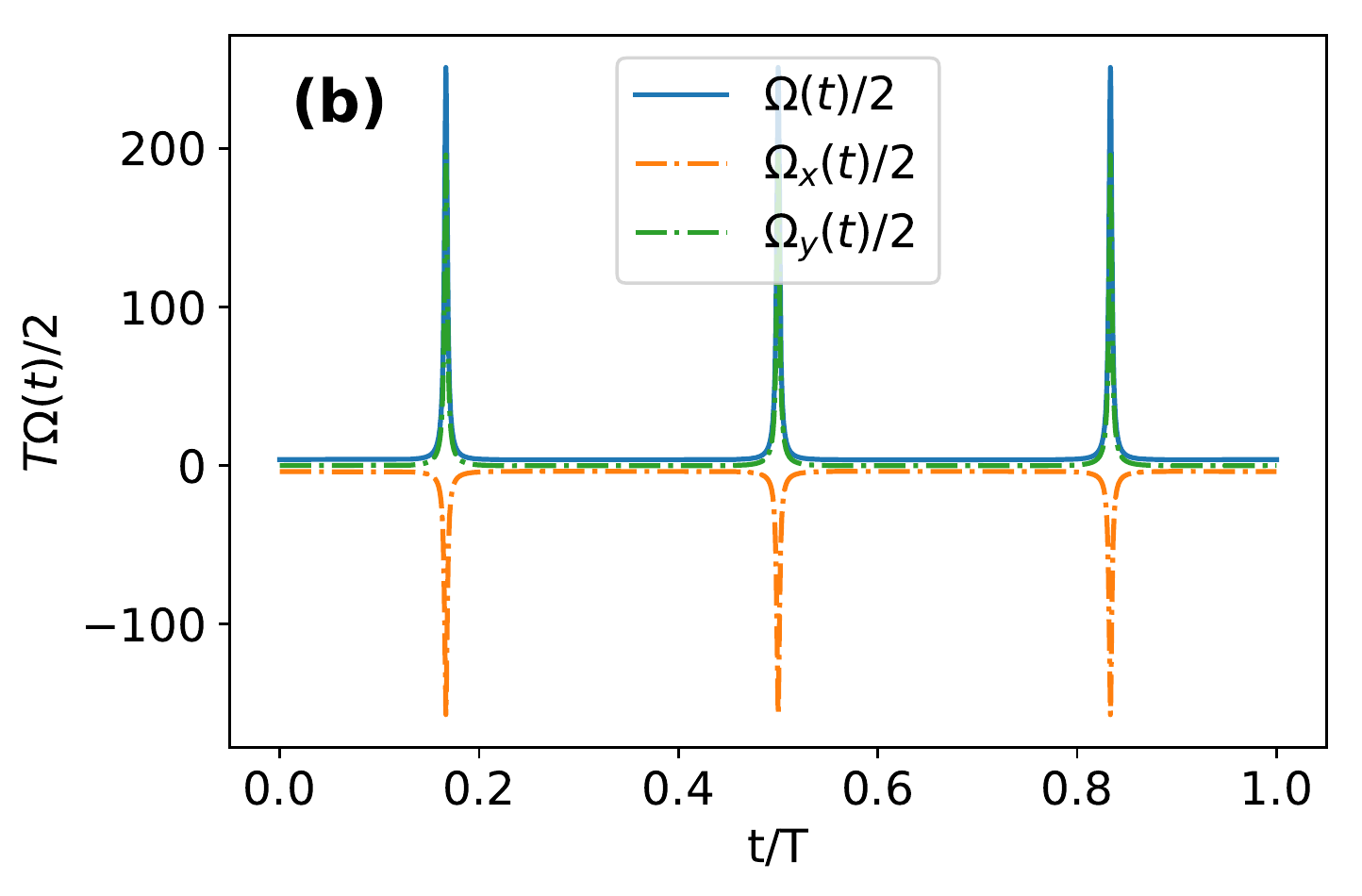}
\caption{Single-qubit identity gate robust against errors up to second order. (a) The curve (blue) and its projections onto the $xy$, $yz$ and $xz$ planes (gray). All three projected curves have zero enclosed area. (b) The pulses obtained from the curvature and torsion of the curve in (a). Here, $\Omega_x=\Omega\cos\Phi$, $\Omega_y=\Omega\sin\Phi$. This figure is adapted from \cite{Zeng_PRA19}.}
\label{fig:curve3}
\end{figure}

In Ref.~\cite{Zeng_PRA19}, it was also shown that second-order errors can be cancelled by designing closed curves with vanishing-area planar projections. This is because the second-order term in the Magnus expansion of $\mathcal{U}_I(t)$ is proportional to
\begin{equation}\label{eq:2nd_order_Magnus}
    \int_0^t\int_0^{t_1}[\mathcal{U}_c^\dag(t_1)\sigma_z\mathcal{U}_c(t_1),\mathcal{U}_c^\dag(t_2)\sigma_z\mathcal{U}_c(t_2)]dt_1dt_2\propto\left(\int_0^t \mathbf{r}(t_1)\times\dot{\mathbf{r}}(t_1) dt_1\right)\cdot\bm{\sigma}.
\end{equation}
Each component of the integral on the right-hand side equals the area enclosed by $\bm{r}(t)$ after it is projected onto a plane orthogonal to the Cartesian unit vector corresponding to that component. Notice that this generalizes a similar result found for plane curves in Sec.~\ref{sec:2d_curves}. Here, we see that in the most general case of driving along two or three axes, three areas must vanish instead of only one. An example of a curve for which all three planar projections vanish in this way is shown in Fig.~\ref{fig:curve3}(a). The pulses extracted from this curve (Fig.~\ref{fig:curve3}(b)) perform an identity operation that is robust to noise up to second order.

\section{Noise-resistant Landau-Zener sweeps through avoided crossings}\label{sec:LandauZener}

In addition to constructing noise-resistant pulses, the SCQC formalism can be used to design high-fidelity control protocols that involve tuning a system close to a noisy avoided crossing. Landau-Zener (LZ) transitions \cite{landau1932,stuckelberg1932,zener1932non,majorana1932} induced by an avoided crossing are widely used for qubit operations, system characterization, initialization, and readout~\cite{shevchenko2010landau,shytov2003landau,ji2003electronic,sun2009population,petta2010coherent,diCarlo2009demonstration,DiCarlo2010preparation,sun2010tunable,Mariantoni2011Implementing,ReedNature2012,cao2013ultrafast,thiele2014electrically,martinis2014fast,wang2018landau,rol2019fast, sillanpaa2006continuous,oliver2005mach,dupont2013coherent,nalbach2013nonequilibrium}. The performance of operations that rely on avoided crossings can be degraded by noise in the energy gap~\cite{huang2011landau}. This can be an issue when driving the system through either a level crossing or an anti-crossing. In the former case, noise fluctuations can create a small energy gap, converting the crossing into an anti-crossing and causing unwanted LZ transitions. Similarly, in situations where the objective is to tune the system to a desired level on the opposite side of an avoided crossing, noise can cause undesirable transitions to the other level. Noise can also lower the fidelity in cases where anti-crossings are exploited to perform operations. Several methods have been developed to address these issues dynamically, including composite LZ pulses~\cite{hicke2006fault}, super-adiabatic LZ pulses~\cite{bason2012high, martinis2014fast}, and LZ sweeps based on geometric phases~\cite{gasparinetti2011geometric,tan2014demonstration,zhang2014realization,wang2016experimental}. However, these methods can sometimes lead to long control times, experimentally impractical pulse waveforms, or imperfect noise cancellation. 

\begin{figure}
\centering
\includegraphics[width=0.5\columnwidth]{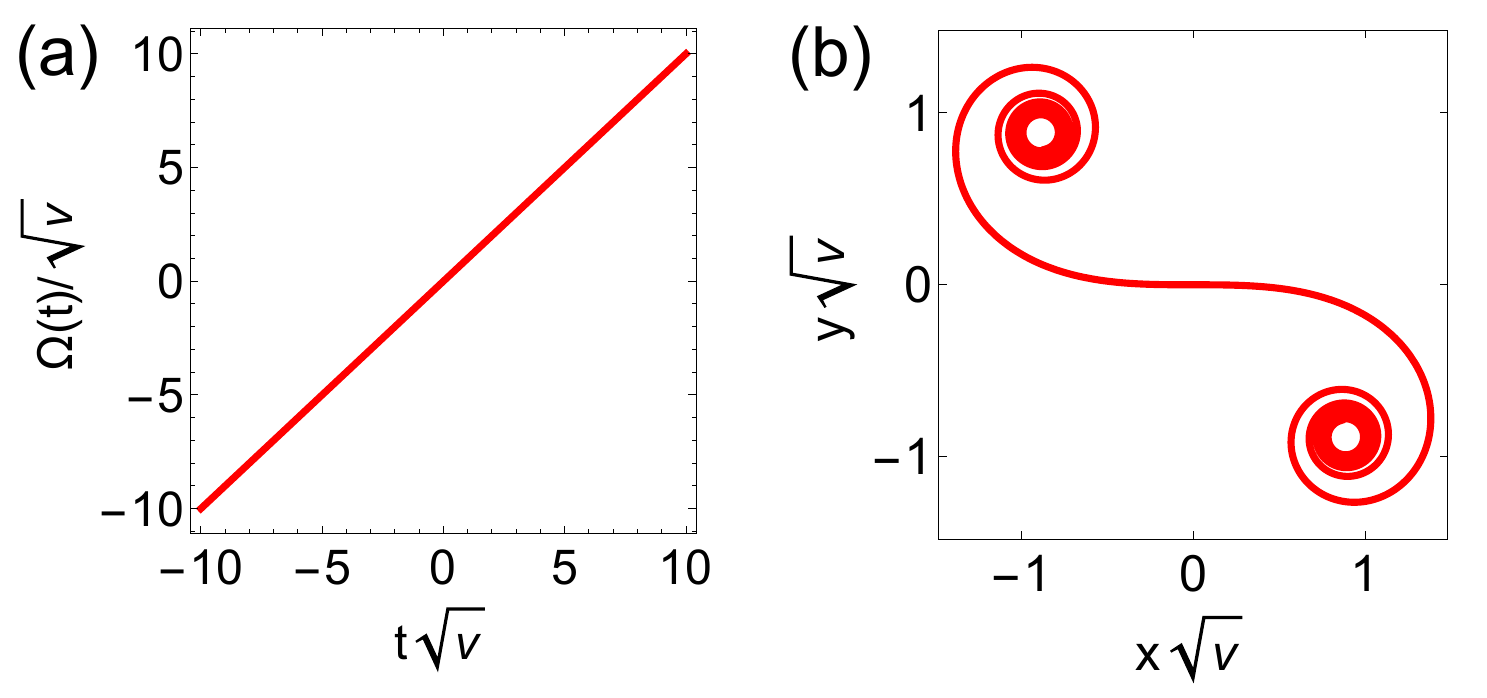}
\caption{A linear Landau-Zener sweep (a) translates to an Euler spiral (b) in the SCQC formalism. This figure is adapted from \cite{zhuang2021noiseresistant}, which has been submitted for publication.}\label{fig:cornu}
\end{figure}

The LZ problem is typically described by a Hamiltonian of the form $\mathcal{H}(t)=\frac{\Omega(t)}{2}\sigma_z+\frac{\Delta}{2}\sigma_x+\epsilon\sigma_x$. $\Delta$ is the energy gap of the anti-crossing at zero bias ($\Omega=0$), while $\epsilon$ is a stochastic fluctuation in this energy. This is similar to the Hamiltonian considered in the previous section (Eq.~(\ref{eq:hamil}) with $\Phi=0$), except that the driving is now along the $z$ axis, while the noise is along $x$. This choice of basis is more natural for the LZ problem, where one typically tunes the system energy to approach the anti-crossing. In this context, we are generally not interested in pulses $\Omega(t)$ that start and end at zero like before, but rather the focus is on control fields with nonzero initial and final values. This can be accounted for in the SCQC formalism by constructing curves that possess nonzero curvature at the initial and final points. For this Hamiltonian, the curvature and torsion are given by $\kappa(t)=-\Omega(t)$ and $\tau=-\Delta$, respectively~\cite{zhuang2021noiseresistant}. In terms of the geometry of space curves, two important differences arise here compared to the previous section. The first is that here we want to allow the curvature to assume both positive and negative values since in the LZ problem, one is generally interested in what happens when the system is swept through the anti-crossing at $\Omega=0$. However, the curvature of a space curve is usually defined to be strictly nonnegative as in Eq.~(\ref{eq:curvature_torsion}). Ref.~\cite{zhuang2021noiseresistant} modified this definition slightly to allow for both positive and negative curvatures. A second important difference compared to the previous section is that here, we are interested in the case of constant $\Delta$, which means that we must find space curves that have constant torsion. A general procedure for carrying out this nontrivial task was presented in Ref.~\cite{zhuang2021noiseresistant}. 

Ref.~\cite{zhuang2021noiseresistant} showed that by carefully designing $\Omega(t)$, it is possible to suppress unwanted transitions caused by the noise fluctuation $\epsilon$, such that the desired state or gate operation is realized at the final time $T$ with high fidelity. We first discuss the case in which the avoided crossing is caused purely by noise, i.e., in the absence of noise it becomes a crossing. A key ingredient is to notice that the original linear LZ sweep, $\Omega(t)\sim t$~\cite{landau1932,stuckelberg1932,zener1932non,majorana1932}, corresponds to a plane curve known as an Euler spiral (see Fig.~\ref{fig:cornu})~\cite{Levien:EECS-2008-111,Bartholdi2012,MatteoOSA2013,LiLSA2018}.

\begin{figure}
\centering
\includegraphics[width=\columnwidth]{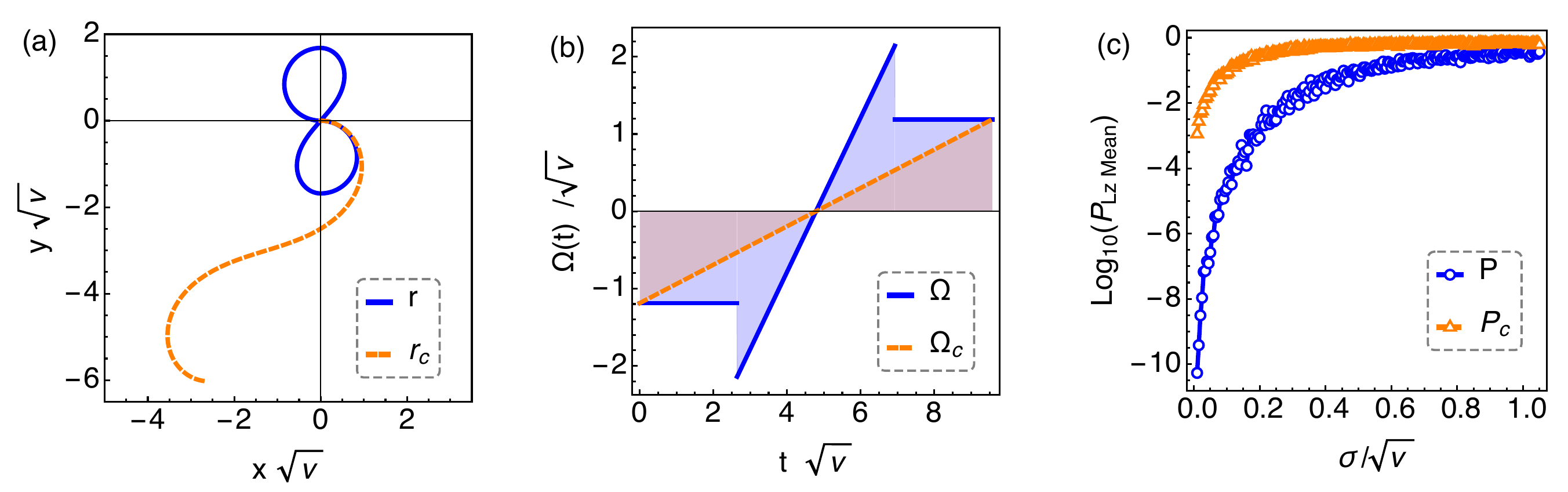}
\caption{(a) A closed curve of zero net area (blue) and an Euler spiral (orange). (b) LZ control fields obtained from the curvatures of the curves in (a). $v$ is the LZ velocity of the linear ramp in the geometrically engineered pulse (blue). (c) LZ probability $P_{\mathrm{LZ}}=|\langle 1|U(T)|0\rangle|^2$ versus noise strength for both pulses shown in (b). For each value of $\sigma$, $P_{\mathrm{LZ}}$ is averaged over 100 instances of $\epsilon$ randomly sampled from a normal distribution with zero mean and standard deviation $\sigma$. This figure is adapted from \cite{zhuang2021noiseresistant}, which has been submitted for publication.}\label{fig:robust_sweep}
\end{figure}

It is evident from the figure that Euler spirals do not close. However, we can combine pieces of Euler spirals together to make closed curves, which in turn yield robust sweep profiles $\Omega(t)$. An example of such a curve constructed from Euler spiral and circular arc segments, along with its corresponding pulse, are shown in Fig.~\ref{fig:robust_sweep}. The figure also shows the probability of an unwanted LZ transition as a function of noise strength. A naive linear sweep is included for comparison, and it is evident that the SCQC-engineered sweep protocol performs better by several orders of magnitude. A striking feature of the robust LZ sweep shown in Fig.~\ref{fig:robust_sweep}(b) is that it is non-monotonic. Ref.~\cite{zhuang2021noiseresistant} in fact proved that it is impossible to cancel noise using a monotonic sweep. This is essentially a consequence of a result in differential geometry known as the Tait-Kneser theorem. Ref.~\cite{zhuang2021noiseresistant} also presented families of arbitrary-angle single-axis gates that are robust to noise up to second order.
\begin{figure}
\centering
\includegraphics[width=0.6\columnwidth]{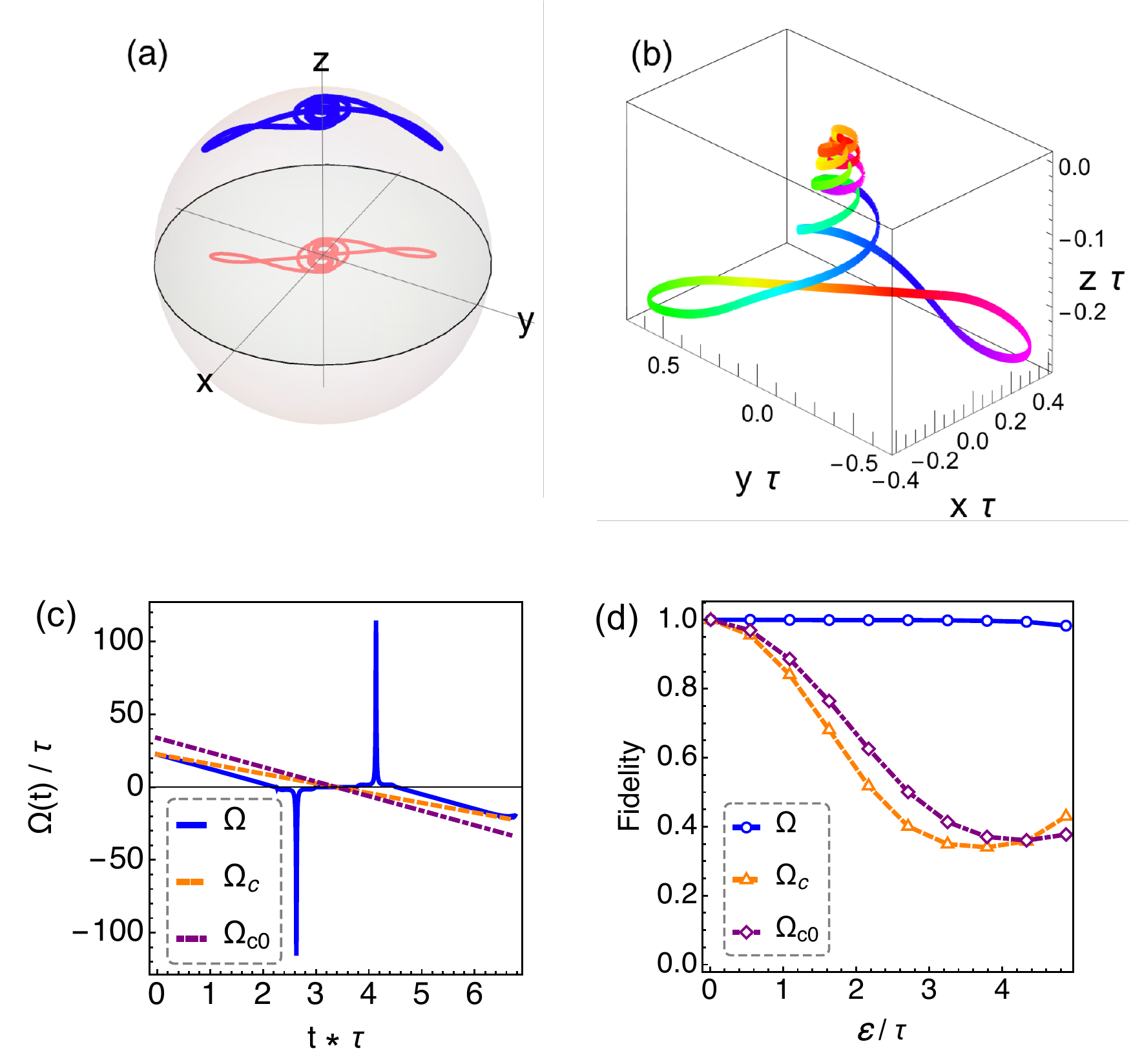}
\caption{A modified Landau-Zener sweep that cancels noise to first order in the case of a nonzero gap $\Delta=|\tau|>0$. (a) The binormal vector $\bm{b}(s)$ obtained by projecting a closed plane curve $\bm{p}(s)$ (which is chosen to have certain rotational symmetries) onto the unit sphere. (b) The space curve $\bm{r}(t)$ corresponding to the binormal curve shown on the left. (c) The pulse (blue) obtained from the curvature of the space curve. Two different linear sweeps (orange and purple) are also shown for comparison. (d) Process fidelity versus noise strength $\epsilon$ (in units of the energy gap $\Delta=|\tau|$) for the SCQC sweep (blue) and naive linear sweeps (orange and purple). This figure is adapted from \cite{zhuang2021noiseresistant}, which has been submitted for publication.}\label{fig:LZspacecurve}
\end{figure}

These ideas can be generalized to the case of a nonzero gap ($\Delta>0$). This requires developing a technique to construct closed curves of constant torsion, which is well known in the differential geometry community to be a challenging problem~\cite{weiner1977closed}. Ref.~\cite{zhuang2021noiseresistant} introduced a general procedure to accomplish this task. One starts by drawing a closed planar curve, $\mathbf{p}(s)=[x(s),y(s)]$, that satisfies certain rotational symmetries. This is then projected onto the surface of a unit sphere to form the binormal indicatrix $\bm{b}(t)$ of a space curve. (After projecting onto the sphere, the binormal must be reparameterized so that $t$ is its arc length.) The space curve can then be obtained from $\bm{b}(t)$ using the formula
\begin{equation}\label{eq:rFromb}
    \mathbf{r}(t)=-\frac{1}{\Delta}\int_0^t\mathbf{b}(s)\times\dot{\mathbf{b}}(s)ds.
\end{equation}
We can see from this formula that closure of the space curve corresponds to three vanishing-area conditions for the three Cartesian projections of $\bm{b}(t)$. This is mathematically similar to the vanishing-area constraints on $\bm{r}(t)$ needed for second-order noise cancellation (see Eq.~(\ref{eq:2nd_order_Magnus})). In the present problem, these vanishing-area constraints can be satisfied by building rotational symmetries into the initial plane curve $\bm{p}(s)$. Once we obtain the space curve from the binormal curve, we can read off the driving field from its curvature. An example is shown in Fig.~\ref{fig:LZspacecurve}. The sweep profile obtained from this example has rather sharp features, but these can be softened using additional curve engineering. The figure also shows the fidelity as a function of noise strength for both the SCQC sweep and two noise-sensitive linear sweeps. It is evident that the former performs substantially better as expected.

\section{Space curve formalism for multi-level and multi-qubit systems}\label{sec:generalized_geometric}

\subsection{Control Hamiltonians from generalized curvatures}

The results described above only apply for individual qubits. It is equally important to find dynamical methods to combat noise during multi-qubit operations, which are a fundamental requirement for most quantum information technologies. It is also necessary to develop such techniques for multi-level systems since logical qubit states always reside in a larger spectrum of energy levels. Achieving high-precision control over the logical states requires taking into account the states outside this subspace, as they can give rise to ``leakage" errors due to population escaping from the logical subspace.

In Ref.~\cite{Buterakos_PRXQ2021}, it was discovered that a geometric structure also underlies the Schr\"odinger equation for multi-level and multi-qubit systems. These more general cases have a geometric interpretation in terms of curves in higher dimensions. A curve in any number of dimensions is described by a set of vectors called the Frenet-Serret basis vectors and by a set of ``generalized curvatures" that generalize the concepts of curvature and torsion that arise in three dimensions, as discussed in Sec.~\ref{sec:non-resonant}. The key to extending the geometric formalism to larger Hilbert spaces is to determine how these basis vectors and curvatures can be derived from a given multi-level or multi-qubit Hamiltonian. A systematic procedure that achieves this will be described after a brief review of the Frenet-Serret basis and generalized curvatures.

A curve $\bm{r}(t)$ in $d$-dimensional Euclidean space defines a set of Frenet-Serret basis vectors $\{\bm{e}_n\}$, $n=1,...,d$. As in the single-qubit case described above, each point along the curve $\bm{r}(t)$ is labeled by the evolution time $t$. For each value of $t$, the Frenet-Serret vectors form an orthonormal basis: $\bm e_m(t)\cdot\bm e_n(t)=\delta_{nm}$. The first vector, $\bm{e}_1(t)$, is chosen to be the tangent vector of the curve at time $t$: $\bm{e}_1(t)=\dot{\bm{r}}(t)$. Thus, the Frenet-Serret frame rigidly rotates with the curve as time progresses. The orthonormality condition immediately implies that $\dot{\bm{e}}_m\cdot\bm{e}_n=-\bm{e}_m\cdot\dot{\bm{e}}_n$. In the case $m=n$, it follows that $\dot{\bm{e}}_n\cdot\bm{e}_n=0$, or in other words $\dot{\bm{e}}_n$ lies in a direction orthogonal to $\bm e_n$. In the case of the first vector, this direction {\it defines} $\bm{e}_2$: $\dot{\bm{e}}_1=\kappa_1\bm{e}_2$, where $\kappa_1$ is the magnitude of $\dot{\bm{e}}_1$. The time derivative of $\bm{e}_2$ can then be written as $\dot{\bm{e}}_2=-\kappa_1\bm{e}_1+\kappa_2\bm{e}_3$, where the first term is included to ensure that $\dot{\bm{e}}_1\cdot\bm{e}_2=-\bm{e}_1\cdot\dot{\bm{e}}_2$ is satisfied, and where $\bm{e}_3$ is defined to be the component of $\dot{\bm{e}}_2$ that is orthogonal to both $\bm{e}_1$ and $\bm{e}_2$. Continuing on to $\dot{\bm{e}}_3$, etc., and following the same logic then leads to definitions of the remaining $\bm{e}_n$, as well as to a set of self-consistency conditions known as the Frenet-Serret equations:
\begin{equation}
\dot{\bm{e}}_n=-\kappa_{n-1}\bm{e}_{n-1}+\kappa_n\bm{e}_{n+1},\label{eq:FS1}
\end{equation}
where the $\kappa_n$ are referred to as generalized curvatures, and $\bm e_0=0=\bm e_{d+1}$. If we return to the special case of $d=3$ dimensions, then we recognize $\kappa_1$ as the usual curvature, while $\kappa_2$ is the torsion.

If quantum evolution under the Schr\"odinger equation can be represented by a geometric space curve, then it should be possible to identify a set of Frenet-Serret vectors for a given Hamiltonian and show that they satisfy Eq.~(\ref{eq:FS1}). This identification will in turn yield relations between the generalized curvatures and control fields in the Hamiltonian.
Consider the following Hamiltonian:
\begin{equation}
\mathcal{H}=\mathcal{H}_c(t)+\delta \mathcal{H},\label{eq:genham}
\end{equation}
where $\mathcal{H}_c$ is the control Hamiltonian, and $\delta \mathcal{H}$ is a noise term. Define $Q\equiv \delta \mathcal{H}/|\delta \mathcal{H}|$, and assume that $\{\mathcal{H}_c,Q\}=0$. Here, the magnitude of an operator is defined by $|{\cal O}|=\sqrt{{\cal O}\cdot{\cal O}}$, where the inner product of two operators is ${\cal O}_1\cdot{\cal O}_2=k^{-1}\hbox{Tr}\left({\cal O}_1{\cal O}_2\right)$, with $k$ the dimension of the Hilbert space. Note that if $Q$ is a Pauli string on $n$ qubits, then we can always transform into a frame in which the condition $\{\mathcal{H}_c,Q\}=0$ holds. In this case, we would also have that $Q^2=\mathbbm{1}$, which we will assume in what follows for the sake of simplicity.
Define a set of unit vectors expressed in terms of a set of operators $\{A_n\}$ with $n=1,...,d$:
\begin{equation}
\bm{e}_n=\mathcal{U}_c^\dagger A_nQ\mathcal{U}_c,\label{FSbasis}
\end{equation}
where $\mathcal{U}_c$ is the evolution operator generated by $\mathcal{H}_c$: $i\dot{\mathcal{U}}_c=\mathcal{H}_c\mathcal{U}_c$. These vectors will satisfy $\bm{e}_n\cdot\bm{e}_n=\mathbbm{1}$ provided $A_n$ either commutes or anti-commutes with $Q$ (and provided $A_n$ is normalized appropriately). Differentiating Eq.~(\ref{FSbasis}), yields
\begin{equation}
\dot{\bm{e}}_n=i\mathcal{U}_c^\dagger\{\mathcal{H}_c,A_n\}Q\mathcal{U}_c+\mathcal{U}_c^\dagger\dot{A}_nQ\mathcal{U}_c.\label{eq:FS2}
\end{equation}
It is tempting to identify the two terms in Eq.~(\ref{eq:FS2}) with the two terms in Eq.~(\ref{eq:FS1}). Ref.~\cite{Buterakos_PRXQ2021} showed that this can in fact be done if the $A_n$ obey the following recursion relation:
\begin{equation}\label{eq:generalized_recursion}
\kappa_nA_{n+1}=\bigg\{\begin{matrix} i\{\mathcal{H}_c,A_n\} & \hbox{if $n$ is odd} \cr \dot A_n & \hbox{if $n$ is even}\end{matrix}\quad,
\end{equation}
where $A_1=\mathbbm{1}$.
The curvatures, $\kappa_n$, can be obtained by taking the magnitudes of these expressions, since $A_n$ has unit magnitude. Eq.~(\ref{eq:generalized_recursion}) thus provides a general mapping between control fields in the Hamiltonian and the generalized curvatures.
From Eq.~(\ref{eq:generalized_recursion}), it can be seen that the $A_n$ obey the following (anti)commutation relations:
\begin{equation}
[A_1,Q]=0,\;\;\{A_2,Q\}=0,\;\;\{A_3,Q\}=0,\;\;[A_4,Q]=0,\;\;[A_5,Q]=0,\;\;\{A_6,Q\}=0,\ldots
\end{equation}
Thus, the normalization of the Frenet-Serret vectors is guaranteed. 

As in the single-qubit case, the components $x_\alpha(t)$ of the space curve $\bm{r}(t)$ are given by the integral of the tangent vector:
\begin{equation}\label{eq:space_curve_general_case}
\sum_{\alpha=1}^d x_\alpha(t)V_\alpha=\int_0^tdt'\bm{e}_1(t')=\int_0^tdt'\mathcal{U}_c^\dag(t')Q\mathcal{U}_c(t').
\end{equation}
The number of distinct basis operators $V_\alpha$ that emerge from the final integral determines the dimension $d$ of the Euclidean space in which the space curve resides. This in turn depends on the number of operators in $\mathcal{H}_c$ and on the commutativity of these operators with each other and with $Q$, the operator governing the error term $\delta\mathcal{H}$. Because the final integral in Eq.~(\ref{eq:space_curve_general_case}) is proportional to the leading-order term in a Magnus expansion of the interaction-picture evolution operator generated by $\mathcal{U}_c^\dag(t)\delta\mathcal{H}\mathcal{U}_c(t)$, we again see that cancellation of the leading-order error in $\mathcal{U}(T)$ requires $\bm{r}(t)$ to be closed: $\bm{r}(T)=\bm{r}(0)=0$. Thus, by constructing closed curves in $d$ dimensions, we can obtain noise-cancelling control fields from the generalized curvatures of these curves.

Relations (\ref{eq:generalized_recursion}) tell us how the generalized curvatures of the space curve are related to the various terms of the Hamiltonian. However, this does not yet constitute a general procedure for obtaining error-correcting control fields for multi-level or multi-qubit systems. A remaining challenge is that not all curves correspond to physical control fields. As a simple example of this, consider the case of a single qubit driven by an off-resonant pulse, as described in Sec.~\ref{sec:non-resonant}. Here, the system is described by a curve in $d=3$ dimensions with constant torsion (corresponding to a constant pulse detuning). However, a general curve in three dimensions has a torsion that varies in time, and so it does not describe a constant-detuning pulse. As described in Sec.~\ref{sec:LandauZener}, this problem was solved in Ref.~\cite{zhuang2021noiseresistant} in the three-dimensional case. In higher dimensions however, a general procedure to systematically find curves with a number of generalized curvatures held constant remains an open problem. Although we do not yet have a general recipe, explicit examples of curves constrained in this way can still be found in special cases, as we show in the next section.

Another important aspect of the multi-level/multi-qubit SCQC formalism that remains to be understood is how to formulate higher-order noise-cancellation constraints. The results described above show that closed curves guarantee leading-order noise cancellation as in the single-qubit case. In the case of a single qubit, Ref.~\cite{Zeng_PRA19} showed that second-order noise errors are also cancelled by ensuring that the planar projections of the curve have vanishing enclosed area (see Fig.~\ref{fig:curve3}). How this condition generalizes in the multi-level/multi-qubit context has not yet been worked out.

\subsection{Noise-cancellation in multi-qubit quantum gates}\label{sec:multi-qubit_noise_cancellation}

In Ref.~\cite{Buterakos_PRXQ2021}, it was shown that the multi-dimensional SCQC formalism described above can be used to design noise-cancelling pulses for an important class of two-qubit problems. This class includes systems in which the two qubits are coupled via an Ising-like interaction, as is the case for superconducting transmon qubits in the dispersive regime \cite{Koch_PRA07,Economou_PRB15,Deng_PRB17,McKay_PRL19,Magesan_PRA20,Collodo_arxiv20} and also for capacitively coupled or exchange-coupled singlet-triplet spin qubits in semiconductor quantum dots~\cite{Shulman_Science12,Wardrop_PRB14,Wang_NPJQI15,Nichol_NPJQI17,CalderonVargas_PRB19,Qiao_arxiv06}.
In these cases, the two-qubit control Hamiltonian has the form
\begin{equation}
\mathcal{H}_c=\left(\begin{matrix} E_1 & \Omega(t) & 0 & 0 \cr \Omega(t) & -E_1 & 0 & 0 \cr 0 & 0 & E_2 & \Omega(t) \cr 0 & 0 & \Omega(t) & -E_2 \end{matrix}\right).\label{eq:two_qubit_H0}
\end{equation}
Here, $\Omega(t)$ is the driving field we wish to design, while $E_1$ and $E_2$ are constant energy splittings. This Hamiltonian describes two coupled qubits (where the coupling creates the difference between $E_1$ and $E_2$) in a situation where only the second qubit is driven. This provides enough control to create two-qubit entanglement~\cite{Economou_PRB15,Deng_PRB17}. We can also use the formalism to derive pulses that rotate only one qubit without changing the state of the other qubit and while also suppressing noise. This is a challenging problem for quantum processors in which the couplings are always on~\cite{McKay_PRL19,Magesan_PRA20}.

\begin{figure}
\centering
\includegraphics[width=0.5\columnwidth]{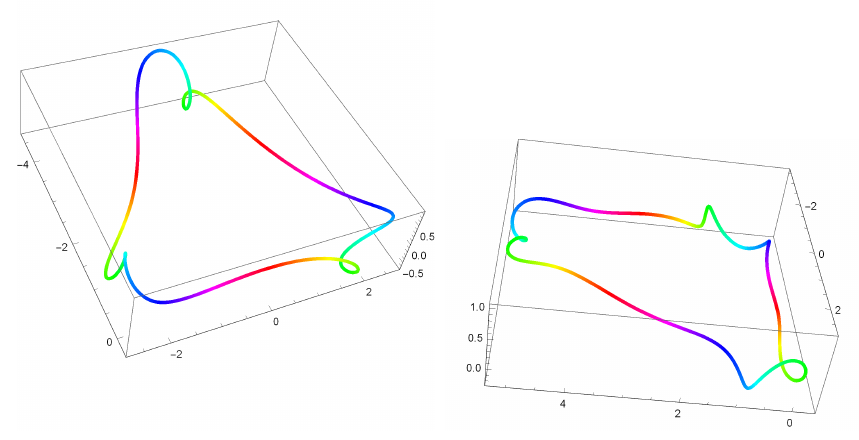}
\includegraphics[width=0.3\columnwidth]{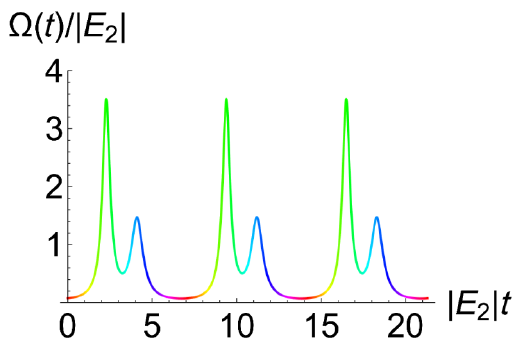}
\caption{Designing pulses for two-qubit gates using the multi-qubit SCQC formalism. Two closed 3-dimensional curves of constant torsion are shown on the left. The curves are colored to illustrate which points correspond to which part of the pulse, $\Omega(t)$, they generate (right). This figure was adapted from \cite{Buterakos_PRXQ2021}.}\label{fig:donovan}
\end{figure}

\begin{figure}
\centering
\includegraphics[width=0.5\columnwidth]{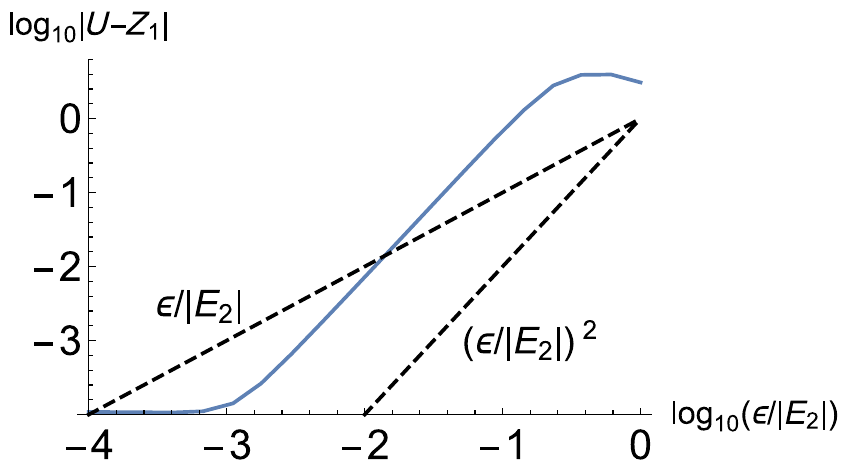}
\caption{Infidelity versus noise strength of the pulse shown in Fig.~\ref{fig:donovan} (solid blue curve).  The dashed lines show linear and quadratic  scaling  with  noise  strength.   It  is  clear  that  the gate infidelity is consistent with quadratic scaling, indicating that first-order noise errors have been canceled. This figure was adapted from \cite{Buterakos_PRXQ2021}.}\label{fig:donovan_fid}
\end{figure}

Ref.~\cite{Buterakos_PRXQ2021} studied the case where the two-qubit system is subject to slow noise fluctuations in the energy levels of both qubits. In the context of superconducting transmon qubits, these fluctuations typically arise from magnetic flux noise~\cite{Krantz_APR19}, while for semiconductor spin qubits they are caused by nuclear spin noise~\cite{Medford_PRL12,Malinowski_PRL17}. These fluctuations are described by the noise Hamiltonian $\delta \mathcal{H}=\epsilon \mathbbm{1}\otimes\sigma_z$. Ref.~\cite{Buterakos_PRXQ2021} showed that the evolution of the two-qubit system can be mapped to a curve in $d=6$ dimensions, with generalized curvatures:
\begin{align}
\kappa_1&=2|\Omega|,\nonumber\\
\kappa_2&=\sqrt{2(E_1^2+E_2^2)},\nonumber\\
\kappa_3&=\frac{\sqrt{2}|E_1^2-E_2^2|}{\sqrt{E_1^2+E_2^2}},\nonumber\\
\kappa_4&=2\sqrt{\Omega^2+\frac{2E_1^2E_2^2}{E_1^2+E_2^2}},\nonumber\\
\kappa_5&=\frac{E_1E_2\sqrt{2(E_1^2+E_2^2)}}{\Omega^2(E_1^2+E_2^2)+2E_1^2E_2^2}\;\frac{d\Omega}{dt}.
\label{eqn:curvatures}
\end{align}
This curve must be closed in order for the leading-order error to cancel.
At first glance, constructing such a curve appears challenging because two of the five curvatures ($\kappa_2$ and $\kappa_3$) must be held constant, while the remaining three depend on a single function, $\Omega(t)$. The six-dimensional curves that are relevant to this problem are thus highly constrained, making it harder to construct examples. However, it was found that this difficulty can be circumvented by decomposing the six-dimensional curves into two curves in three dimensions. The fact that this can be done can be seen from Eq.~(\ref{eq:two_qubit_H0}), where the $4\times4$ Hamiltonian contains two $2\times2$ blocks. Each block can be viewed as an effective two-level system and can thus be mapped to a three-dimensional space curve. Both two-level systems have the same driving field $\Omega(t)$ but different ``detunings": $2E_1$ in the first block and $2E_2$ in the second. Therefore, to obtain physical solutions, one must construct two space curves that have the same curvature and distinct but constant torsions. Ref.~\cite{Buterakos_PRXQ2021} constructed examples of such curves by piecing together constant-torsion segments to form closed loops. An example pair of such curves that satisfy all the necessary constraints and the corresponding pulse are shown in Fig.~\ref{fig:donovan}. The resulting pulse implements a $z$ rotation on one qubit while leaving the other qubit alone. The fact that this pulse successfully suppresses noise at the same time is demonstrated in Fig.~\ref{fig:donovan_fid}, which shows the gate infidelity as a function of the noise strength. Ref.~\cite{Buterakos_PRXQ2021} also constructed an example of a pair of curves that yield a maximally entangling CNOT gate. In this case, one of the two curves must exhibit a cusp at the origin, so that an $x$ rotation is implemented in one of the $2\times2$ subspaces.

This example illustrates that decomposing a higher-dimensional curve into lower-dimensional curves can make it easier to obey physical constraints. However, it remains to develop a systematic procedure that can be applied to more general Hamiltonians, particularly ones that do not exhibit a block-diagonal structure like in Eq.~(\ref{eq:two_qubit_H0}), as well as to Hamiltonians that contain more than one driving field.

\section{Cancelling time-dependent noise errors}\label{sec:time-dependent}

In the works described in the previous sections, it was assumed the noise is quasistatic (i.e., $\epsilon$ in Eqs.~(\ref{eq:Hamresdriving}) and (\ref{eq:hamil}) is constant over the duration of the pulse $\Omega(t)$). This is a good first approximation for most qubit platforms, including superconducting transmon qubits, trapped ions, and semiconductor spin qubits. However, in reality the noise possesses a slow time-dependence, and this can become important if the goal is to achieve very high control precision, as is necessary to reach quantum error-correction thresholds~\cite{Fowler_PRA12}. This motivates extending the SCQC formalism to the case of time-dependent noise. This was done in the case of single-qubit gates in Ref.~\cite{bikun}.

Consider the general scenario in which a qubit couples to a bath with many degrees of freedom, so that the Hamiltonian is
	\begin{equation}\label{eq:Hamiltonian0}
	    H(t) = \frac{\Omega(t)}{2}\sigma_x\otimes \mathbbm{1}
	    + \lambda \sigma_z \otimes B
	    + \mathbbm{1}\otimes H_B,
	\end{equation}
	where $\Omega(t)$ is the control field and $\sigma_\alpha\;(\alpha = x,y,z)$ are Pauli matrices acting on the qubit. The operators $B$ and $H_B$ only act on the environmental degrees of freedom and represent a generic quantum bath. The second term is the qubit-bath interaction with coupling strength $\lambda$. This interaction induces pure dephasing and is responsible for decohering the qubit. The goal is to design control pulses that implement quantum gates while dynamically decoupling the qubit from the bath. If the bath degrees of freedom fluctuate slowly compared to the gate time, then this problem can be recast in terms of a stochastic noise parameter $\epsilon$ as in Refs.~\cite{Zeng_NJP2018,Zeng_PRA18,Zeng_PRA19}. However, if the bath fluctuates during the gate, then this will give rise to additional errors that cannot be eliminated using the methods of Refs.~\cite{Zeng_NJP2018,Zeng_PRA18,Zeng_PRA19}. These errors are quantified by the leading-order gate infidelity~\cite{bikun}:
	\begin{equation}\label{eq:infidelity_leading}
	    1-\mathcal{F} \approx 
	    \frac{1}{3}\int_{-\infty}^\infty\frac{d\omega}{2\pi} S(\omega)F(\omega,T),
	\end{equation}
where $S(\omega)$ is the noise power spectrum (the Fourier transform of the two-point correlation function $\langle B(t)B(t')\rangle$ of the bath operators), which quantifies how much noise is present at frequency $\omega$. $F(\omega,T)\equiv |f(\omega,T)|^2 + |f(-\omega,T)|^2$ is called a filter function, where
	\begin{equation}\label{eq:fdef}
	    f(\omega,T)\equiv \int_0^T dt e^{i[\phi(t)-\omega t]},
	\end{equation}
	and $\phi(\tau)=\int_0^\tau dt \Omega(t)$ is the integral of the pulse. The similarity of this expression for $f(\omega,T)$ to the noise-cancellation constraints in the case of quasistatic noise, Eq.~(\ref{eq:curve_constraints}), suggests that there is an underlying geometric framework in the case of time-dependent noise too. This is indeed the case, as will now be shown.

The leading time-dependent noise error will be suppressed if the integral in Eq.~(\ref{eq:infidelity_leading}) is small. This in turn requires $F(\omega,T)$ to be small whenever $S(\omega)$ is large. The task then is to find pulses $\Omega(t)$ so that $F(\omega,T)$ exhibits this property. In most solid-state qubit platforms, the noise is concentrated at low frequencies (which is why the quasistatic approximation is effective at describing the bulk of the error). This means that we need to make the filter function as flat as possible in the vicinity of $\omega=0$, or in other words, not only do we need $f(0,T)=0$, but also the first $k-1$ derivatives should vanish: 
\begin{equation}\label{eq:derivative_constraints}
\frac{d^\ell}{d(\omega T)^\ell}f(\omega,T)\bigg|_{\omega=0}=0,\qquad \ell=0,1,\ldots,k-1.
\end{equation}

\begin{figure}[h]
    	\centering
    	\includegraphics[width=0.8\linewidth]{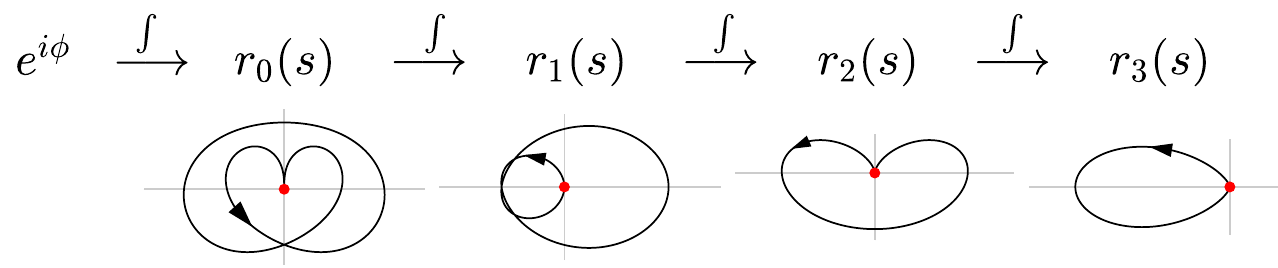}
    	\caption{An example of a sequence of closed curves $r_\ell(t)=d^{3-\ell}r_3/dt^{3-\ell}$, where $k=4$. The red dots represent the origin, which is also the starting/ending point of the curves. In this example, $r_0$ is the integral of $e^{i\phi(t)}$. The integral of $r_3$ is no longer a closed curve, so the sequence terminates. This figure is adapted from \cite{bikun}, which has been submitted for publication.
    	}
    	\label{fig:rmsequence}
    \end{figure}

\begin{figure}[h]
    	\centering
    	\includegraphics[width=0.6\linewidth]{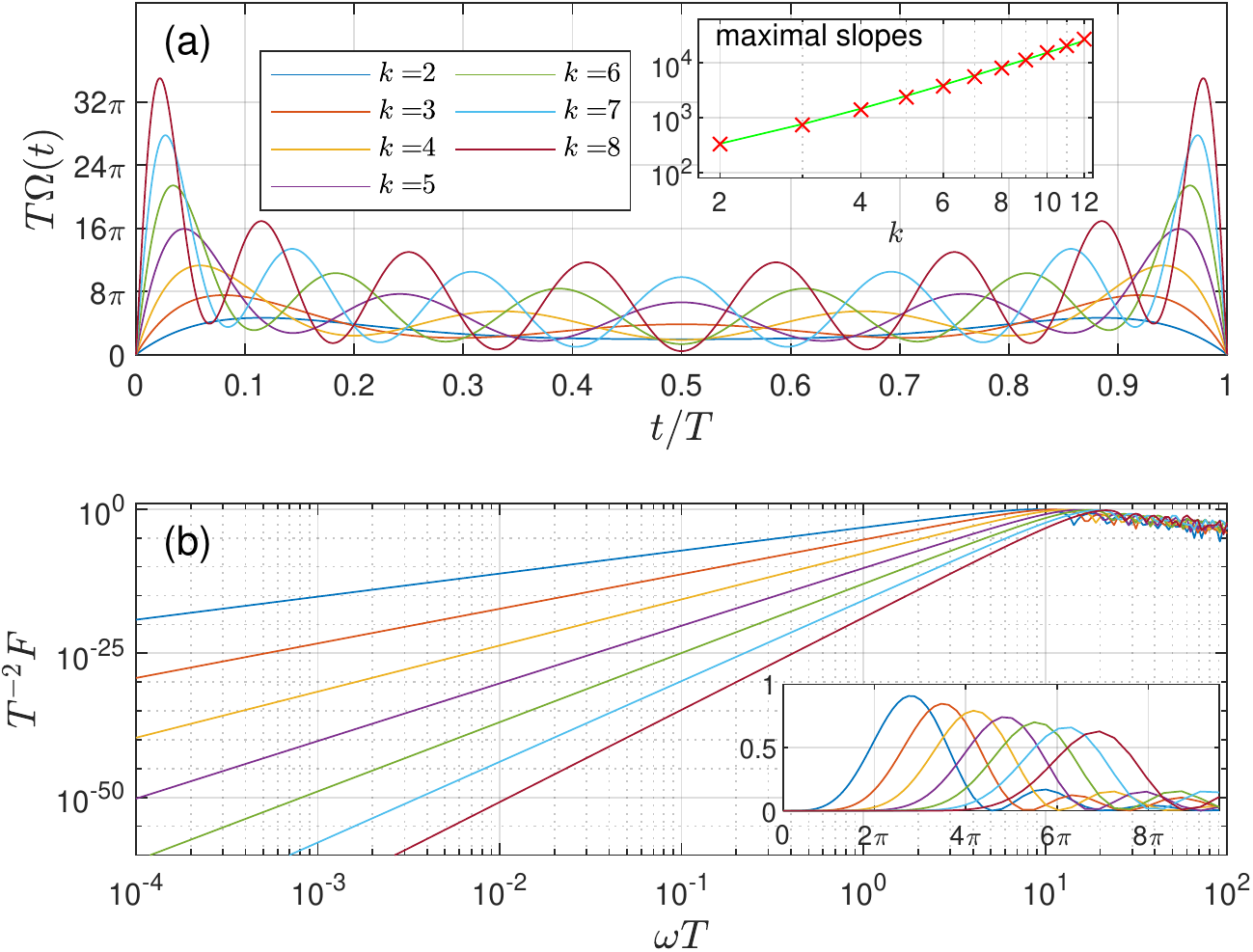}
    	\caption{(a) Seven pulses (colored lines) which eliminate $k$ derivatives of the filter function for $k=2,...,8$. Each pulse implements a logical NOT gate on a single qubit. The inset shows the maximum slopes (pulse bandwidth) required for different $k$, which has an approximate $\sim k^3$ growth at small $k$ (green line).
    	(b) The corresponding filter functions (colored lines), whose asymptotic slopes $\approx 2k$ indicate the $\mathcal{O}(\omega^{2k})$ suppression near $\omega T=0$. 
    	The inset shows the same filter function in a linear plot. This figure is adapted from \cite{bikun}, which has been submitted for publication.}
    	\label{fig:Omegak2to8}
    \end{figure}

There is a geometrical interpretation of these constraints. If we define the following series of plane curves,
\begin{equation}
r_m(t) = \int_0^{t}d{t_1}\int_0^{t_1}d{t_2}\cdots\int_0^{t_{m}}d{t_{m+1}} e^{i\phi(t_{m+1})},
\end{equation}
labelled by integer $m$,
then through repeated integrations by parts it follows that
\begin{equation}
	\begin{aligned}
    i^\ell\frac{ d^\ell}{d(\omega T)^\ell} [T^{-1}f( \omega,T)]_{\omega T=0}
    = 
	    \int_0^T s^\ell e^{i\phi(t)}d{t} =\sum_{m=0}^\ell \frac{(-1)^m \ell !}{(\ell - m)!} r_m(T),
	\end{aligned}
\end{equation}
for $\ell=0,\ldots,k-1$. Thus, the first $k$ derivatives of $f(\omega,T)$ will vanish if each of the $r_m(t)$ for $m=0,1,\cdots,k-1$ corresponds to a closed curve. Therefore, suppressing time-dependent noise errors requires that we construct a sequence of closed curves where each curve in the sequence is the integral of the previous. An example of such a sequence is illustrated in Fig.~\ref{fig:rmsequence}. Ref.~\cite{bikun} developed a procedure for generating curve sequences like this numerically using an iterative damped Newton method. The corresponding pulse can be obtained from the curvature of the curve at the bottom of the hierarchy, $r_0(t)=d^{k-1}r_{k-1}/dt^{k-1}$. Examples of pulses obtained in this way for values of $k$ ranging from 2 to 8 are shown in Fig.~\ref{fig:Omegak2to8}, along with the resulting filter functions. The increasing flatness of the filter function near $\omega=0$ with increasing $k$ is evident in the figure. This behavior is similar to that given by the UDD dynamical decoupling sequence~\cite{Uhrig_PRL07}.

A number of qubit platforms suffer from colored noise such as $1/f$ noise. This type of noise originates from nuclear spin noise and charge noise in the case of semiconductor spin qubits~\cite{Medford_PRL12,Dial_PRL13,Malinowski_PRL17} and from magnetic flux noise~\cite{Krantz_APR19} in the case of superconducting qubits. In both cases, it is the dominant source of qubit dephasing. An important question concerns the tradeoff between pulse bandwidth and the size of the frequency window in which noise is cancelled. From Fig.~\ref{fig:Omegak2to8}, it is evident that suppressing noise over larger frequency ranges requires the use of pulses that contain higher frequency components. At some point, the pulse bandwidth will exceed what is possible to implement experimentally due to waveform generator limitations. Note that this is a generic feature of cancelling time-dependent noise and is not specific to the SCQC approach. However, it is likely that some pulses will achieve the same noise suppression while using less bandwidth compared to others.
It is also important to note that, as indicated by a no-go theorem presented in Ref.~\cite{Liu_PRA_Nogo_softcutoff}, the noise filtration suggested by Eq.~(\ref{eq:infidelity_leading}) is only efficient for higher $k$ if $S(\omega)$ has a hard frequency cutoff. This makes the suppression of $1/f$ noise particularly challenging. Nonetheless, it was found in Ref.~\cite{bikun} that the smooth pulses shown in Fig.~\ref{fig:Omegak2to8} can still outperform delta-function sequences such as UDD~\cite{Uhrig_PRL07} to some extent, due to their always-on nature. 

\section{Simultaneous cancellation of multiple noise sources}\label{sec:multiple_noises}

In addition to noise errors that act transversely to the drive, it is often the case that errors enter into the driving field itself. For a single qubit, the Hamiltonian can then take the form
\begin{equation}\label{eq:two_noise_ham}
    \mathcal{H}(t)=\frac{\Omega(t)+\delta\Omega(t)}{2}\sigma_x+\frac{\Delta}{2}\sigma_z+\epsilon\sigma_z.
\end{equation}
The error in the driving field is represented by the stochastic fluctuation $\delta\Omega(t)$. This can be caused by systematic errors that arise in the process of generating and sending the pulse to the device, or it can be due to a second source of environmental noise (on top of the source that causes the transverse fluctuation $\epsilon$). An example of the latter is charge noise in singlet-triplet spin qubits, which can cause time-dependent fluctuations in the exchange coupling used to drive qubit rotations~\cite{Martins_PRL16,Reed_PRL2016}. The question is then whether it is possible to find pulses $\Omega(t)$ that dynamically correct both error terms simultaneously. Approaches based on group theory or composite pulses have been developed to address this problem in general settings~\cite{Viola_PRL03,Brown_PRA2004,Khodjasteh_PRL2009,Green_NJP13,Merrill_Wiley14}. More recent methods based on extensions of the SCQC framework have also been developed, including a direct extension in which additional pulse constraints are imposed to ensure the cancellation of pulse-amplitude errors~\cite{Throckmorton2019}, a geometric technique based on enforcing noise-cancellation constraints on the tantrix~\cite{Barnes_SciRep15,Gungordu2019}, and a method known as ``doubly geometric gates" that combines holonomic evolution with space curves~\cite{dong2021doubly}. Here, we review each of these geometric methods in turn and discuss their relative merits.

\subsection{Closed space curves with pulse amplitude constraints}

Ref.~\cite{Throckmorton2019} considered a Hamiltonian of the form (\ref{eq:two_noise_ham}) but with $\Delta=0$ and where the pulse-amplitude noise fluctuation is of the form $\delta\Omega(t)=f[\Omega(t)]\eta$, where $f[\Omega(t)]$ is a function of the pulse amplitude, and $\eta$ is a small stochastic constant. We can expand the evolution operator in powers of $\epsilon$ and $\eta$. The first-order term in $\epsilon$ can be used to map the qubit evolution to a plane curve as in Sec.~\ref{sec:2d_curves}, and requiring this term to vanish again leads to the closed-curve condition. The requirement that the first-order term in $\eta$ vanishes at the final time $T$ yields an additional non-local constraint on the curvature, $\Omega(t)$, of the plane curve:
\begin{equation}\label{eq:amplitude_constraint}
    \int_0^Tdtf[\Omega(t)]=0.
\end{equation}
If the noise causes a stochastic rescaling of the pulse amplitude, then this is described by setting $f[\Omega(t)]=\Omega(t)$, in which case Eq.~(\ref{eq:amplitude_constraint}) requires the area of the pulse to vanish. This means that we can only dynamically correct identity operations in this case. It is also immediately clear from Eq.~(\ref{eq:amplitude_constraint}) that if the noise creates a random pulse offset described by $f[\Omega(t)]=1$, then it is impossible to cancel noise for any gate operation. On the other hand, Ref.~\cite{Throckmorton2019} showed that for $f[\Omega(t)]\sim\Omega(t)^k$, it is possible to cancel the leading-order noise in both $\epsilon$ and $\eta$ while performing nontrivial gates for odd values of $k$ other than 1. Explicit examples using sequences of four square or trapezoidal pulses were given for several values of $k$. Even values of $k$ cannot be canceled because the integral in Eq.~(\ref{eq:amplitude_constraint}) is then strictly positive for any $T>0$. 

It is important to note that the inability to cancel pulse rescaling errors ($k=1$) while performing arbitrary $x$ rotations is a consequence of having set $\Delta=0$. For $\Delta\ne0$, it was shown in Ref.~\cite{Kestner_PRL13} that it is possible to cancel both types of noise errors at the same time while executing arbitrary single-qubit gates using composite {\sc supcode} pulses. Next, we discuss two geometric approaches that supplement closed space curves with additional conditions that suppress pulse-amplitude noise when $\Delta\ne0$. These approaches allow for arbitrary control waveforms instead of relying on fixed pulse shapes.

\subsection{Reverse-engineering approach}

The first geometric approach to designing gates that dynamically correct against two types of noise simultaneously that we discuss here was presented in Ref.~\cite{Barnes_SciRep15}. Rather than starting from space curves, this approach begins from a method for reverse-engineering the evolution of a qubit that was introduced in Refs.~\cite{Barnes_PRL12,Barnes_PRA13}. The goal of these earlier works was to circumvent the analytical intractability of the general time-dependent two-level Schr\"odinger equation, which complicates the process of designing quantum logic gates even in the absence of noise. These works developed a systematic procedure to find control fields for which the resulting evolution operator can be obtained analytically; the key idea was to utilize an auxiliary function $\chi(t)$ from which both the evolution operator and the control fields can be determined. It was shown that for a Hamiltonian of the form (\ref{eq:two_noise_ham}) (in the absence of noise: $\epsilon=0$, $\delta\Omega=0$), any $\chi(t)$ that satisfies the constraint $|\dot\chi|\le|\Delta|$ yields an analytical solution for the evolution operator along with the pulse $\Omega(t)$ that generates it. Leveraging this result, Ref.~\cite{Barnes_SciRep15} then considered the case where both noise terms are present ($\epsilon\ne0$, $\delta\Omega\ne0$) and derived constraints that ensure that the leading-order errors in the evolution operator vanish at the final time:
\begin{equation}
    \mathcal{U}(T)=\mathcal{U}_c(T)+\mathcal{O}(\epsilon^2,\delta\Omega^2,\epsilon\delta\Omega).
\end{equation}
It was further shown that the noise cancellation constraints can be interpreted as constraints on the shape of a curve that lives on the surface of a sphere. If we parameterize the evolution operator as in Eq.~(\ref{eq:evol_parameterized}), then this sphere is parameterized by $\theta$ and $\zeta$. Although it was not realized in Ref.~\cite{Barnes_SciRep15}, this curve is in fact the tantrix $\dot{\bm{r}}(t)$ of a three-dimensional space curve. Moreover, the highly non-local constraints for cancelling the first-order term in $\epsilon$ derived in that work are precisely the closed curve condition described in Sec.~\ref{sec:non-resonant}. If one works with the tantrix, then this condition becomes a set of integral constraints that can be challenging to satisfy. On the other hand, the constraints for cancelling pulse-amplitude noise found in Ref.~\cite{Barnes_SciRep15} are not easily understood from the point of view of space curves. Instead of using the auxiliary function $\chi(t)$, these constraints can alternatively be derived by considering a first-order Magnus expansion in the pulse error $\delta\Omega$, which then leads to the noise-cancellation conditions
\begin{equation}
    \int_0^T\delta\Omega(t')\mathcal{U}_c^\dag(t')\sigma_x\mathcal{U}_c(t')=0.
\end{equation}
Using the relation between $\mathcal{U}_c(t)$ and the tantrix, Eq.~(\ref{eq:tantrix}), this can be recast as a highly non-local constraint on the space curve. Given a form for the pulse fluctuation, e.g., $\delta\Omega(t)=\Omega(t)\eta$ where $\eta\ll1$, this constraint can be solved by starting from an ansatz for the space curve containing several adjustable parameters that can be varied without altering the closed curve condition needed to cancel $\epsilon$ noise. It can prove quite challenging to find smooth pulses using this approach, although there has been recent progress in simplifying the noise-cancellation conditions to facilitate the process of finding solutions~\cite{Gungordu2019}.

\subsection{Doubly geometric gates}\label{sec:DoGs}

\begin{figure} 
    \centering
    \includegraphics[width=0.5\textwidth]{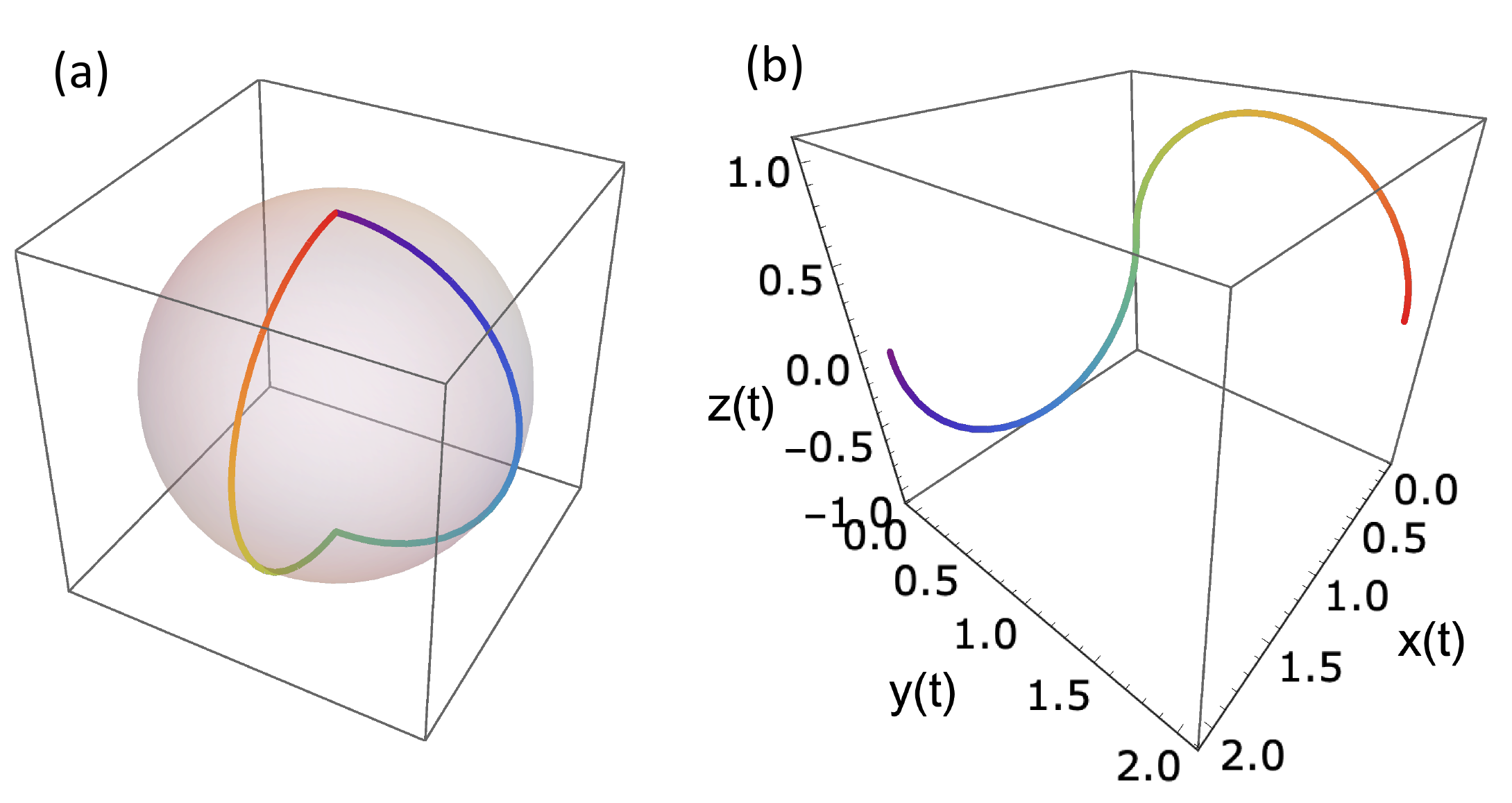}
    \caption{(a) Bloch sphere representation and (b) space curve $\bm{r}(t)$ of the standard ``orange-slice"  holonomic gate, which is specified by two geodesic lines that differ by a $\pi/2$ azimuthal angle. The holonomic trajectory on the Bloch sphere encloses a $\pi/2$ geometric phase.  These curves are colored to indicate which part of the Bloch sphere trajectory corresponds to which part of the space curve, starting from red at $t=0$ and ending with violet at $t=T$. This figure is adapted from \cite{dong2021doubly}, which has been submitted for publication.}
    \label{fig:Wenzheng_DoG_fig1}
\end{figure}

An alternative approach to suppressing control field errors is to use holonomic gates~\cite{ZANARDI199994}. Here, the idea is to base gate operations on geometric phases~\cite{Berry.Geometric.Phase,ANANDAN1988171,Wilczek.Zee,Berry_2009,SolinasPRA2004,Sjoqvist.IJQC.2015,Xu.PRL.2012,Utkan.JPSJ.2014}. Because these phases are given by the solid angle enclosed by the evolution path traced on the Bloch sphere (Fig.~\ref{fig:Wenzheng_DoG_fig1}(a)), such gates are insensitive to noise errors that leave this path invariant. This includes pulse-amplitude errors that alter the rate at which the Bloch sphere trajectory is traversed while leaving its shape intact. Although geometric phases were originally defined in the context of adiabatic evolution~\cite{Berry.Geometric.Phase}, this concept was later generalized to non-adiabatic evolution~\cite{PhysRevLett.58.1593}, enabling fast holonomic gates~\cite{Sj_qvist_2012,Sjoqvist,XueZhengyuan.PRA.2018,Ribeiro.PRA.2019,LiuBoajie.PRL.2019,YingZuJian.PRR.2020,Shkolnikov.PRB.2020,li2020dynamically,Ji2021noncyclic,Zhao.holonomicDD.PRL.2021}. Such gates have been experimentally implemented in a number of qubit platforms~ \cite{YanTongxing.PRL.2019,SunLuYan.PRL.2020,Duan.Science.2001,AiMingzhong.PRApp.2020,PhysRevLett.119.140503,Zhou2016,Sekiguchi.2017}. However, such gates generally remain sensitive to transverse noise errors such as the $\epsilon$ term in Eq.~(\ref{eq:two_noise_ham}). In Ref.~\cite{dong2021doubly}, it was shown that this issue can be addressed by combining space curves with holonomic evolution to construct ``doubly geometric" (DoG) gates that simultaneously suppress two orthogonal noise sources. In this section, we review how this approach works.

A single-qubit holonomic gate is constructed by starting from a state that evolves cyclically under a given control Hamiltonian, which can in general be taken to have the form given in Eq.~(\ref{eq:hamil}). Without loss of generality, we can set the initial state of this cyclic evolution to be the computational state $\ket{0}$. Using the parameterization from Eq.~(\ref{eq:evol_parameterized}) (but now for the full evolution operator $\mathcal{U}_c$ instead of $\widetilde{\mathcal{U}}_c$) and setting $\zeta=2\alpha+\phi+\pi/2$, $\lambda=-\phi-\pi/2$, the state at later times becomes
\begin{equation}
    \ket{\psi(t)}=e^{i\alpha(t)}\left[\cos(\theta(t)/2)\ket{0}+\sin(\theta(t)/2)e^{i\phi(t)}\ket{1}\right].
\end{equation}
The evolution is then cyclic if $\theta(T)=\theta(0)=0$, in which case the evolution trajectory on the Bloch sphere starts and ends at the north pole. The net phase $\alpha(T)$ accumulated in the process is purely geometric if the evolution satisfies the parallel transport condition: $\dot{\alpha}(t)=-\frac{1}{2}\big(1-\cos\theta(t)\big)\dot{\phi}(t)$. A fully-specified holonomic evolution of state $\ket{\psi(t)}$ is sufficient to determine a holonomic quantum gate~\cite{dong2021doubly}. A holonomic $z$-rotation by angle $\alpha(T)$ is then implemented, where $\alpha(T)=-\frac{1}{2}\int^T_0(1-\cos\theta)d\phi$ is the enclosed solid angle. Given $\theta(t)$ and $\phi(t)$, the three control fields $\Omega(t)$, $\Phi(t)$, $\Delta(t)$ that generate this evolution can be obtained from $\mathcal{H}_c=i\dot{\mathcal{U}}_c\mathcal{U}_c^\dag$.

Ref.~\cite{dong2021doubly} mapped common examples of holonomic evolutions onto space curves and showed that these curves are not closed in general, indicating a failure to suppress $\epsilon$ noise. One such example is given in Fig.~\ref{fig:Wenzheng_DoG_fig1}, which shows the open space curve corresponding to the well-known ``orange-slice model" holonomic gate~\cite{Mousolou.PRA.2014,SunLuYan.PRL.2020,AiMingzhong.PRApp.2020}. 
The fact that these curves do not close is expected since such noise typically acts transversely to the Bloch sphere trajectory. However, Ref.~\cite{dong2021doubly} went on to demonstrate that it is possible to find holonomic evolutions for which the space curve is closed. Moreover, a general recipe for finding such evolutions was presented. In particular, it was shown that there is a unique holonomic evolution associated with every smooth, closed space curve and that the control fields that generate this evolution can be obtained from the space curve using the following expressions:
\begin{equation}\label{eq:holonomic_fields}
        \Omega(t)=\|\ddot{\bm{r}}(t)\|,\qquad
        \Delta(t)=\frac{\dot x\ddot y-\dot y\ddot x}{\dot z},\qquad
        \Phi(t)=\int_0^t[\tau(t')+\Delta(t')]dt'.
\end{equation}
Before employing these expressions, it is important to first ensure that the curve $\bm{r}(t)$ has been parameterized such that $\|\dot{\bm{r}}(t)\|=1$. Any smooth, closed space curve can therefore be used to construct DoG gates that are simultaneously holonomic and robust to $\epsilon$ noise. It is interesting to note that while a given space curve corresponds to a family of control Hamiltonians related by frame transformations (as noted in Sec.~\ref{sec:non-resonant}), the parallel transport condition picks out a unique Hamiltonian from this family.

\begin{figure}
    \centering
    \includegraphics[width=0.6\columnwidth]{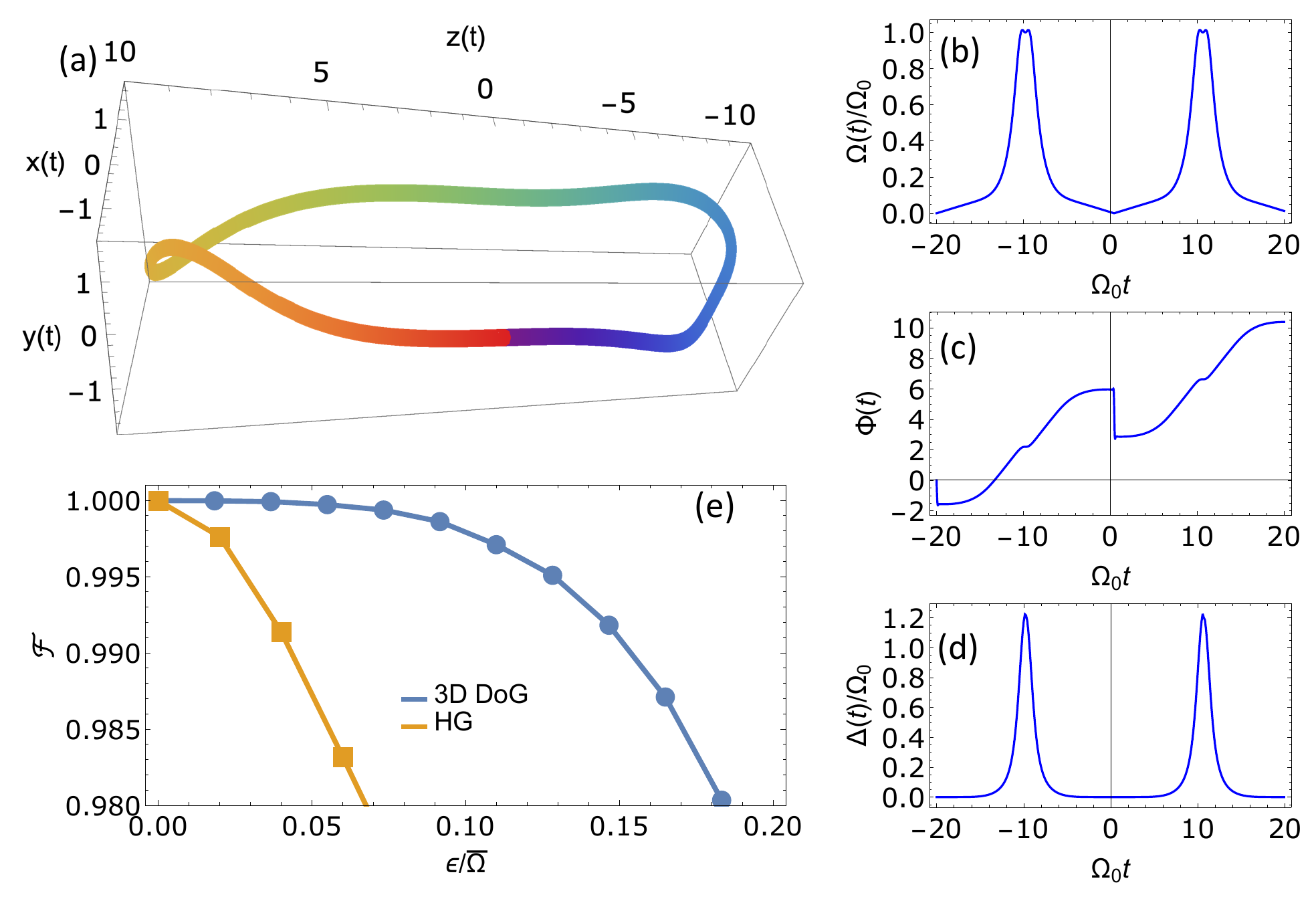}
    \caption{(a) Twisted space curve $\bm{r}_\xi(t)$ for $\xi=\pi/2000$. (b-d) DoG control fields $\Omega(t)$, $\Phi(t)$, $\Delta(t)$ generated from $\bm{r}_\xi(t)$ in (a). (e) Gate fidelities of the DoG gate constructed from  $\bm{r}_\xi(t)$ in (a) versus detuning error rate (the ratio of the detuning error $\epsilon$ to the time-averaged driving strength $\bar{\Omega}$). Results for the standard orange-slice model-based holonomic gate (HG) are shown for comparison. This figure is adapted from \cite{dong2021doubly}, which has been submitted for publication.}
    \label{fig:Wenzheng_DoG_fig2}
\end{figure}

Ref.~\cite{dong2021doubly} presented several explicit examples of DoG gates. A continuous family of such gates with adjustable geometric phases $\alpha(T)$ was obtained by starting from a plane curve $\bm{r}_0(t)=y(t)\hat y+z(t)\hat z$ and applying an operation that twists the curve into the third dimension ($x$). The twisted version of the curve is obtained from the formula
\begin{equation}
    \widetilde{\bm{r}}_\xi(t)= -(y-\pi/2)\sin(\xi z^3)\hat x+(y-\pi/2)\cos(\xi z^3)\hat y+z\hat z,
\end{equation}
where $\xi$ is the ``twist" parameter. Here, the original planar curve $\bm{r}_0(t)$ is chosen such that its curvature is given by a superposition of hyperbolic secant pulses: $\Omega(t)=\Omega_0\hbox{sech}(\Omega_0t-10)\Theta(20-\Omega_0t)+\Omega_0\hbox{sech}(\Omega_0t-30)\Theta(\Omega_0t-20)$, where $\Theta$ represents the step function, and the final time is $T=40/\Omega_0$. The motivation for starting from hyperbolic secant pulses is that they have nice analytical properties that facilitates their use in experiments~\cite{PhysRevB.74.205415,EconomouPRL2007,EconomouPRB2012,Greilich2009,KuPRA2017}. After twisting, it is necessary to reparameterize the curve to obtain a new curve $\bm{r}_\xi(t)$ such that the tantrix is properly normalized: $\|\dot{\bm{r}}_\xi(t)\|=1$. For each value of $\xi$, we can then employ Eq.~(\ref{eq:holonomic_fields}) to obtain the control fields that implement the DoG gate. As an example, the curve and resulting control fields for $\xi=\pi/2000$ are shown in Fig.~\ref{fig:Wenzheng_DoG_fig2}(a-d); the corresponding holonomic gate is $U(\xi=\pi/2000)= \text{diag}\{e^{-i0.41\pi},e^{i0.41\pi}\}$.  
The noise-cancelling power of this DoG gate is confirmed in Fig.~\ref{fig:Wenzheng_DoG_fig2}(e), which shows the gate fidelity as a function of the noise error. For comparison, the fidelity for the  orange-slice model holonomic gate is also shown. Both types of gate are chosen to implement the same operation $U(\xi=\pi/2000)$, which is a $z$-rotation whose angle depends linearly on $\xi$.  It is clear that the DoG gate substantially outperforms the standard orange-slice gate in the presence of this type of error.

\section{Outlook}\label{sec:outlook}

\begin{figure}
\centering
\includegraphics[width=\columnwidth]{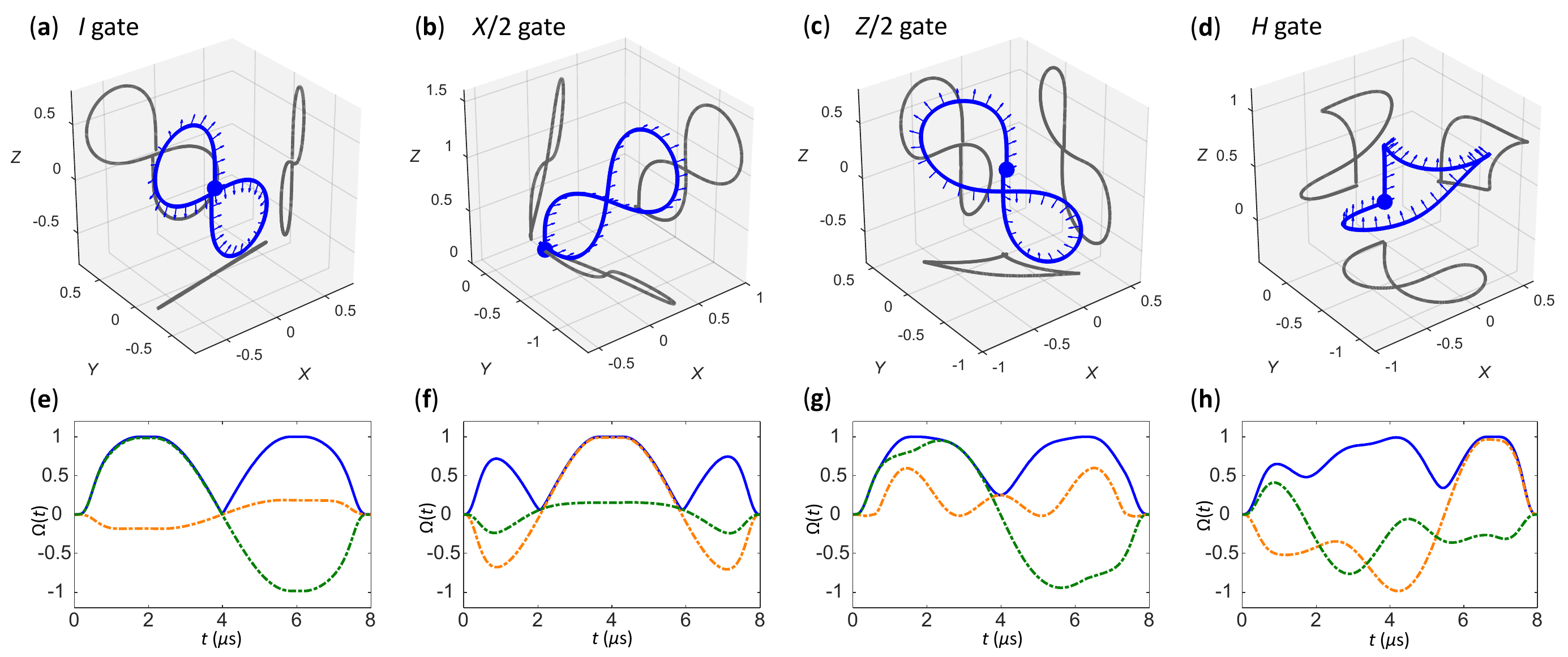}
\caption{Using space curves to analyze pulses obtained from GRAPE. (a-d) The space curves and their projections onto the $xy$, $yz$, and $xz$ planes corresponding to four different microwave pulses (e-h) designed to implement four different single-qubit gates (identity, $\pi/2$ rotation about $x$ ($X/2$), $\pi/2$ rotation about $z$ ($Z/2$), and Hadamard gate ($H$)) while cancelling noise in a silicon quantum dot spin qubit~\cite{Yang_NatElec19}. Arrows on the curves represent the phase in the evolution operator that is controlled by the total torsion (e.g., Eq.~(\ref{eq:evol_phase})). For example, for the $I$ gate in (a) this phase is 0, while for the $Z/2$ gate in (c) it is $\pi/2$. In panels (e-h), the dashed orange and green curves are $\Omega_x(t)/2$ and $\Omega_y(t)/2$, while the solid blue curve is the total magnitude of the pulse envelope. This figure is adapted from \cite{Zeng_PRA19}.}
\label{fig:grapes}
\end{figure}

The results reviewed here show that the SCQC formalism provides access to the full solution space of noise-resistant control fields. Because all such control fields correspond to closed space curves, we can identify the control Hamiltonians that generate globally optimal gate operations by looking for closed curves that obey certain constraints dictated by the physics of the system and other experimental considerations. As discussed at various points above, there remain several directions in need of further investigation, especially in the case of the multi-qubit/multi-level SCQC formalism, such as finding ways to systematically construct curves in higher dimensions with constant generalized curvatures, including multiple (possibly time-dependent) noise sources, and finding the shortest curves under a given set of constraints to obtain time-optimal pulses for a given task.

In addition to these future directions, it is also important to emphasize that SCQC is complementary to state-of-the-art numerical methods for designing pulses~\cite{konnov1999global,palao2002quantum,Khaneja_2005,brif2010control,doria2011optimal,caneva2011chopped,glaser2015training,suter2016colloquium,van2017robust,Tian_PRA2020}. This is because the SCQC framework provides a global view of the optimal control landscape---something that is rarely possible with numerical techniques. Numerical methods often operate based on local information in the parameter space of control fields, and this can sometimes lead to catastrophic failures that are difficult to predict or diagnose. There are at least two ways in which SCQC could be used as a tool to assist with numerical optimal control techniques. The first is that it could be used to identify good initial guesses that can then be further refined with numerical algorithms. The second is that SCQC can be used as a diagnostic tool to analyze the extent to which numerically generated control fields are successful in cancelling different components of the noise error.

The first steps toward this second application were taken in Ref.~\cite{Zeng_PRA19} by using the geometric framework as a tool to diagnose the noise-cancelling properties of pulses produced by the numerical algorithm known as Gradient Ascent Pulse Engineering (GRAPE)~\cite{Khaneja_2005}. Ref.~\cite{Zeng_PRA19} used the SCQC framework to analyze pulses that were recently designed to implement high-fidelity single-qubit gates on silicon quantum dot spin qubits using GRAPE~\cite{Yang_NatElec19}. Fig.~\ref{fig:grapes} shows the space curves for four such pulses, which perform four different single-qubit gates, including an identity operation ($I$), a $\pi/2$ rotation about $x$ ($X/2$), a $\pi/2$ rotation about $z$ ($Z/2$), and a Hadamard operation ($H$). The GRAPE algorithm is implemented with gate fidelity as the cost function and with a noise level corresponding to $\sqrt{\langle\epsilon^2\rangle}=16.7$ kHz, which was attributed to nuclear spin noise in Ref.~\cite{Yang_NatElec19}. Constraints were also imposed on the pulse bandwidth through filtering, where the pulses are strongly smoothed out and forced to approach zero at the beginning and end of the gate. 

From the figure, it is evident that in each case, the corresponding space curve is almost closed, showing that the first-order error-cancellation constraint is almost perfectly satisfied. Moreover, the two-dimensional projections of the curves form symmetric figure-eight shapes in most cases, showing that the second-order cancellation constraint is nearly satisfied as well. Interestingly, it was found that these pulses needed to be 4-5 times longer than the typical time scale of a $\pi$ pulse (1.75 $\mu$s for the parameters used in Ref.~\cite{Yang_NatElec19}); the reason for this is apparent from the space curve, where the bandwidth constraints require pulse durations on the order of 8 $\mu$s in order for the planar projections of the curves to complete their respective figure-eights and thus suppress second-order noise. It is clear from these results that experimental limitations on pulse amplitude or bandwidth are fully compatible with the SCQC formalism, and that realistic pulses correspond to smooth curves that respect the geometrical noise-cancellation conditions.

A goal of future work will be to combine both geometric and numerical methods to achieve the best performance possible in leading quantum computing platforms. Combining these techniques can make it easier to find control pulses that suppress noise while respecting experimental constraints and while implementing a desired quantum operation as quickly as possible. 

\section*{Acknowledgments}

This  work  is  supported  by  the U.S. Office of Naval Research (N00014-17-1-2971), the U.S. Army  Research Office (W911NF-17-0287), and the U.S. Department of Energy (DE-SC0018326).

\providecommand{\newblock}{}


\begin{thebibliography}{100}
\expandafter\ifx\csname url\endcsname\relax
  \def\url#1{{\tt #1}}\fi
\expandafter\ifx\csname urlprefix\endcsname\relax\def\urlprefix{URL }\fi
\providecommand{\eprint}[2][]{\url{#2}}
\bibliographystyle{iopart-num}

\bibitem{GisinRMP2002}
Gisin N, Ribordy G, Tittel W and Zbinden H 2002 {\em Rev. Mod. Phys.\/} {\bf
  74}(1) 145--195
  \urlprefix\url{https://link.aps.org/doi/10.1103/RevModPhys.74.145}

\bibitem{Gisin_NatPhoton2007}
Gisin N and Thew R 2007 {\em Nature Photonics\/} {\bf 1} 165--171
  \urlprefix\url{https://doi.org/10.1038/nphoton.2007.22}

\bibitem{Ladd_Nature2010}
Ladd T~D, Jelezko F, Laflamme R, Nakamura Y, Monroe C and O'Brien J~L 2010 {\em
  Nature\/} {\bf 464} 45--53
  \urlprefix\url{https://doi.org/10.1038/nature08812}

\bibitem{DevoretScience2013}
Devoret M~H and Schoelkopf R~J 2013 {\em Science\/} {\bf 339} 1169--1174 ISSN
  0036-8075
  \urlprefix\url{https://science.sciencemag.org/content/339/6124/1169}

\bibitem{DegenRMP2017}
Degen C~L, Reinhard F and Cappellaro P 2017 {\em Rev. Mod. Phys.\/} {\bf 89}(3)
  035002 \urlprefix\url{https://link.aps.org/doi/10.1103/RevModPhys.89.035002}

\bibitem{Chirolli_AIP2008}
Chirolli L and Burkard G 2008 {\em Advances in Physics\/} {\bf 57} 225--285
  \urlprefix\url{https://doi.org/10.1080/00018730802218067}

\bibitem{Hahn_PR50}
Hahn E~L 1950 {\em Phys. Rev.\/} {\bf 80} 580

\bibitem{Carr_Purcell}
Carr H~Y and Purcell E~M 1954 {\em Phys. Rev.\/} {\bf 94} 640

\bibitem{Meiboom_Gill}
Meiboom S and Gill D 1958 {\em Rev. Sci. Instrum.\/} {\bf 29} 688

\bibitem{Haeberlen}
Haeberlen U 1976 {\em High Resolution NMR in Solids, Advances in Magnetic
  Resonance Series, Supplement 1\/} (New York: Academic)

\bibitem{Vandersypen_RMP05}
Vandersypen L~M~K and Chuang I~L 2005 {\em Rev. Mod. Phys.\/} {\bf 76}
  1037--1069

\bibitem{Bluhm_NP11}
Bluhm H, Foletti S, Neder I, Rudner M, Mahalu D, Umansky V and Yacoby A 2011
  {\em Nat. Phys.\/} {\bf 7} 109

\bibitem{Tyryshkin_NatMat11}
Tyryshkin A~M, Tojo S, Morton J~J~L, Riemann H, Abrosimov N~V, Becker P, Pohl
  H~J, Schenkel T, Thewalt M~L~W, Itoh K~M and Lyon S~A 2011 {\em Nat.
  Mater.\/} {\bf 11} 143

\bibitem{Poletto_PRL12}
Poletto S, Gambetta J~M, Merkel S~T, Smolin J~A, Chow J~M, C\'orcoles A~D,
  Keefe G~A, Rothwell M~B, Rozen J~R, Abraham D~W, Rigetti C and Steffen M 2012
  {\em Phys. Rev. Lett.\/} {\bf 109}(24) 240505

\bibitem{Muhonen_NatNano14}
Muhonen J~T, Dehollain J~P, Laucht A, Hudson F~E, Sekiguchi T, Itoh K~M,
  Jamieson D~N, McCallum J~C, Dzurak A~S and Morello A 2014 {\em Nat.
  Nanotechnol.\/} {\bf 9} 986

\bibitem{Malinowski_NatNano17}
Malinowski F~K, Martins F, Nissen P~D, Barnes E, Cywi{\'n}ski {\L}, Rudner M~S,
  Fallahi S, Gardner G~C, Manfra M~J, Marcus C~M and Kuemmeth F 2017 {\em
  Nature Nanotechnology\/} {\bf 12} 16--20
  \urlprefix\url{https://doi.org/10.1038/nnano.2016.170}

\bibitem{Viola_PRA98}
Viola L and Lloyd S 1998 {\em Phys. Rev. A\/} {\bf 58} 2733

\bibitem{Khodjasteh_PRL05}
Khodjasteh K and Lidar D~A 2005 {\em Phys.\ Rev.\ Lett.\/} {\bf 95} 180501

\bibitem{Uhrig_PRL07}
Uhrig G~S 2007 {\em Phys.\ Rev.\ Lett.\/} {\bf 98} 100504

\bibitem{Zhang_Viola_PRB08}
Zhang W, Konstantinidis N~P, Dobrovitski V~V, Harmon B~N, Santos L~F and Viola
  L 2008 {\em Phys.\ Rev.\ B\/} {\bf 77} 125336

\bibitem{Jones_NJP12}
Jones N~C, Ladd T~D and Fong B~H 2012 {\em New\ J.\ Phys.\/} {\bf 14} 093045

\bibitem{Levitt_1986}
Levitt M~H 1986 {\em Progress in Nuclear Magnetic Resonance Spectroscopy\/}
  {\bf 18} 61--122
  \urlprefix\url{https://doi.org/10.1016/0079-6565%2886%2980005-x}

\bibitem{Goelman_JMR89}
Goelman G, Vega S and Zax D~B 1989 {\em J.\ Magn.\ Reson.\/} {\bf 81} 423
  \urlprefix\url{https://www.sciencedirect.com/science/article/pii/0022236489900772}

\bibitem{Wimperis1994}
Wimperis S 1994 {\em J. Magn. Reson. Ser. A\/} {\bf 109} 221--231 ISSN 10641858
  \urlprefix\url{https://www.sciencedirect.com/science/article/abs/pii/S1064185884711594}

\bibitem{Cummins_PRA03}
Cummins H~K, Llewellyn G and Jones J~A 2003 {\em Phys. Rev. A\/} {\bf 67}
  042308

\bibitem{Biercuk_Nature09}
Biercuk M~J, Uys H, VanDevender A~P, Shiga N, Itano W~M and Bollinger J~J 2009
  {\em Nature\/} {\bf 458} 996
  \urlprefix\url{https://www.nature.com/articles/nature07951}

\bibitem{Khodjasteh_PRL10}
Khodjasteh K, Lidar D~A and Viola L 2010 {\em Phys.\ Rev.\ Lett.\/} {\bf 104}
  090501

\bibitem{Jones2010}
Jones J~A 2011 {\em Prog. Nucl. Magn. Reson. Spectrosc.\/} {\bf 59} 91--120
  ISSN 00796565
  \urlprefix\url{https://www.sciencedirect.com/science/article/abs/pii/S0079656510001111}

\bibitem{vanderSar_Nature12}
van~der Sar T, Wang Z~H, Blok M~S, Bernien H, Taminiau T~H, Toyli D, Lidar D~A,
  Awschalom D~D, Hanson R and Dobrovitski V~V 2012 {\em Nature\/} {\bf 484}
  82--86 \urlprefix\url{https://www.nature.com/articles/nature10900}

\bibitem{Wang_NatComm12}
Wang X, Bishop L~S, Kestner J~P, Barnes E, Sun K and {Das Sarma} S 2012 {\em
  Nat. Commun.\/} {\bf 3} 997

\bibitem{Green_NJP13}
Green T~J, Sastrawan J, Uys H and Biercuk M~J 2013 {\em New Journal of
  Physics\/} {\bf 15} 095004
  \urlprefix\url{http://stacks.iop.org/1367-2630/15/i=9/a=095004}

\bibitem{Kestner_PRL13}
Kestner J~P, Wang X, Bishop L~S, Barnes E and {Das Sarma} S 2013 {\em Phys.\
  Rev.\ Lett.\/} {\bf 110} 140502

\bibitem{Calderon-Vargas2016}
Calderon-Vargas F~A and Kestner J~P 2017 {\em Phys. Rev. Lett.\/} {\bf 118}
  150502 ISSN 0031-9007
  \urlprefix\url{http://link.aps.org/doi/10.1103/PhysRevLett.118.150502}

\bibitem{Petta_Science05}
Petta J~R, Johnson A~C, Taylor J~M, Laird E~A, Yacoby A, Lukin M~D, Marcus C~M,
  Hanson M~P and Gossard A~C 2005 {\em Science\/} {\bf 309} 2180--2184

\bibitem{Wang_PRB14}
Wang X, Calderon-Vargas F~A, Rana M~S, Kestner J~P, Barnes E and Das~Sarma S
  2014 {\em Phys. Rev. B\/} {\bf 90}(15) 155306

\bibitem{Rong.14}
Rong X, Geng J, Wang Z, Zhang Q, Ju C, Shi F, Duan C~K and Du J 2014 {\em Phys.
  Rev. Lett.\/} {\bf 112} 050503

\bibitem{Dial_PRL13}
Dial O~E, Shulman M~D, Harvey S~P, Bluhm H, Umansky V and Yacoby A 2013 {\em
  Phys. Rev. Lett.\/} {\bf 110}(14) 146804

\bibitem{OMalley_PRApplied15}
O'Malley P~J~J, Kelly J, Barends R, Campbell B, Chen Y, Chen Z, Chiaro B,
  Dunsworth A, Fowler A~G, Hoi I~C, Jeffrey E, Megrant A, Mutus J, Neill C,
  Quintana C, Roushan P, Sank D, Vainsencher A, Wenner J, White T~C, Korotkov
  A~N, Cleland A~N and Martinis J~M 2015 {\em Phys. Rev. Applied\/} {\bf 3}(4)
  044009
  \urlprefix\url{https://link.aps.org/doi/10.1103/PhysRevApplied.3.044009}

\bibitem{Martins_PRL16}
Martins F, Malinowski F~K, Nissen P~D, Barnes E, Fallahi S, Gardner G~C, Manfra
  M~J, Marcus C~M and Kuemmeth F 2015 {\em Phys.\ Rev.\ Lett.\/} {\bf 116}
  116801

\bibitem{Hutchings_PRApplied17}
Hutchings M~D, Hertzberg J~B, Liu Y, Bronn N~T, Keefe G~A, Brink M, Chow J~M
  and Plourde B~L~T 2017 {\em Phys. Rev. Applied\/} {\bf 8}(4) 044003
  \urlprefix\url{https://link.aps.org/doi/10.1103/PhysRevApplied.8.044003}

\bibitem{Clerk2010}
Clerk A~A, Girvin S~M, Marquardt F, Schoelkopf R~J, Devoret M~H, Girvin S~M,
  Marquardt F and Schoelkopf R~J 2010 {\em Rev. Mod. Phys.\/} {\bf 82}
  1155--1208 ISSN 0034-6861
  \urlprefix\url{https://link.aps.org/doi/10.1103/RevModPhys.82.1155}

\bibitem{Paladino2014}
Paladino E, Galperin Y~M, Falci G and Altshuler B~L 2013 {\em Rev. Mod.
  Phys.\/} {\bf 86} 361--418 ISSN 0034-6861
  \urlprefix\url{https://link.aps.org/doi/10.1103/RevModPhys.86.361
  http://dx.doi.org/10.1103/RevModPhys.86.361}

\bibitem{Schreier_PRB08}
Schreier J~A, Houck A~A, Koch J, Schuster D~I, Johnson B~R, Chow J~M, Gambetta
  J~M, Majer J, Frunzio L, Devoret M~H, Girvin S~M and Schoelkopf R~J 2008 {\em
  Phys. Rev. B\/} {\bf 77}(18) 180502

\bibitem{Reilly_PRL08}
Reilly D~J, Taylor J~M, Laird E~A, Petta J~R, Marcus C~M, Hanson M~P and
  Gossard A~C 2008 {\em Phys.\ Rev.\ Lett.\/} {\bf 101} 236803

\bibitem{Medford_PRL12}
Medford J, Cywi\ifmmode~\acute{n}\else \'{n}\fi{}ski L, Barthel C, Marcus C~M,
  Hanson M~P and Gossard A~C 2012 {\em Phys. Rev. Lett.\/} {\bf 108}(8) 086802

\bibitem{Sank_PRL12}
Sank D, Barends R, Bialczak R~C, Chen Y, Kelly J, Lenander M, Lucero E,
  Mariantoni M, Megrant A, Neeley M, O'Malley P~J~J, Vainsencher A, Wang H,
  Wenner J, White T~C, Yamamoto T, Yin Y, Cleland A~N and Martinis J~M 2012
  {\em Phys. Rev. Lett.\/} {\bf 109}(6) 067001

\bibitem{Anton_PRB12}
Anton S~M, M\"uller C, Birenbaum J~S, O'Kelley S~R, Fefferman A~D, Golubev D~S,
  Hilton G~C, Cho H~M, Irwin K~D, Wellstood F~C, Sch\"on G, Shnirman A and
  Clarke J 2012 {\em Phys. Rev. B\/} {\bf 85}(22) 224505

\bibitem{Sigillito_PRL15}
Sigillito A~J, Tyryshkin A~M and Lyon S~A 2015 {\em Phys. Rev. Lett.\/} {\bf
  114}(21) 217601

\bibitem{Kalra_RSI16}
Kalra R, Laucht A, Dehollain J~P, Bar D, Freer S, Simmons S, Muhonen J~T and
  Morello A 2016 {\em Review of Scientific Instruments\/} {\bf 87} 073905

\bibitem{Zeng_NJP2018}
Zeng J, Deng X~H, Russo A and Barnes E 2018 {\em New J. of Phys\/} {\bf 20}
  033011 \urlprefix\url{https://doi.org/10.1088\%2F1367-2630\%2Faaafe9}

\bibitem{deng2021correcting}
Deng X~H, Hai Y~J, Li J~N and Song Y 2021 Correcting correlated errors for
  quantum gates in multi-qubit systems using smooth pulse control
  (\textit{Preprint} \eprint{2103.08169})

\bibitem{Zeng_PRA18}
Zeng J and Barnes E 2018 {\em Phys. Rev. A\/} {\bf 98}(1) 012301
  \urlprefix\url{https://link.aps.org/doi/10.1103/PhysRevA.98.012301}

\bibitem{Stefanatos_PRA2019}
Stefanatos D and Paspalakis E 2019 {\em Phys. Rev. A\/} {\bf 100}(1) 012111
  \urlprefix\url{https://link.aps.org/doi/10.1103/PhysRevA.100.012111}

\bibitem{Ansel_2021}
Ansel Q, Glaser S~J and Sugny D 2021 {\em Journal of Physics A: Mathematical
  and Theoretical\/} {\bf 54} 085204
  \urlprefix\url{https://doi.org/10.1088/1751-8121/abdba1}

\bibitem{Mandelstam_JPhys45}
Mandelstam L and Tamm I 1945 {\em J. Phys. (USSR)\/} {\bf 9} 249

\bibitem{Margolus_PD98}
Margolus N and Levitin L~B 1998 {\em Physica D\/} {\bf 120} 188

\bibitem{Deffner_PRL13}
Deffner S and Lutz E 2013 {\em Phys. Rev. Lett.\/} {\bf 111}(1) 010402
  \urlprefix\url{https://link.aps.org/doi/10.1103/PhysRevLett.111.010402}

\bibitem{Deffner_2017}
Deffner S and Campbell S 2017 {\em Journal of Physics A: Mathematical and
  Theoretical\/} {\bf 50} 453001
  \urlprefix\url{https://doi.org/10.1088/1751-8121/aa86c6}

\bibitem{Zeng_PRA19}
Zeng J, Yang C~H, Dzurak A~S and Barnes E 2019 {\em Phys. Rev. A\/} {\bf 99}(5)
  052321 \urlprefix\url{https://link.aps.org/doi/10.1103/PhysRevA.99.052321}

\bibitem{landau1932}
Landau L~D 1932 {\em Phys. Z. Sowjetunion\/} {\bf 2} 46

\bibitem{stuckelberg1932}
Stueckelberg E~C~G 1932 {\em Helv. Phys. Acta\/} {\bf 5} 369

\bibitem{zener1932non}
Zener C 1932 {\em Proceedings of the Royal Society of London. Series A,
  Containing Papers of a Mathematical and Physical Character\/} {\bf 137}
  696--702

\bibitem{majorana1932}
Majorana E 1932 {\em Il Nuovo Cimento (1924-1942)\/} {\bf 9} 43--50
  \urlprefix\url{https://doi.org/10.1007/BF02960953}

\bibitem{shevchenko2010landau}
Shevchenko S~N, Ashhab S and Nori F 2010 {\em Physics Reports\/} {\bf 492}
  1--30

\bibitem{shytov2003landau}
Shytov A, Ivanov D and Feigel,AMan M 2003 {\em The European Physical Journal
  B-Condensed Matter and Complex Systems\/} {\bf 36} 263--269

\bibitem{ji2003electronic}
Ji Y, Chung Y, Sprinzak D, Heiblum M, Mahalu D and Shtrikman H 2003 {\em
  Nature\/} {\bf 422} 415--418

\bibitem{sun2009population}
Sun G, Wen X, Wang Y, Cong S, Chen J, Kang L, Xu W, Yu Y, Han S and Wu P 2009
  {\em Applied Physics Letters\/} {\bf 94} 102502

\bibitem{petta2010coherent}
Petta J, Lu H and Gossard A 2010 {\em Science\/} {\bf 327} 669--672

\bibitem{diCarlo2009demonstration}
DiCarlo L, Chow J~M, Gambetta J~M, Bishop L~S, Johnson B~R, Schuster D~I, Majer
  J, Blais A, Frunzio L, Girvin S~M and Schoelkopf R~J 2009 {\em Nature\/} {\bf
  460} 240--244 \urlprefix\url{https://doi.org/10.1038/nature08121}

\bibitem{DiCarlo2010preparation}
DiCarlo L, Reed M~D, Sun L, Johnson B~R, Chow J~M, Gambetta J~M, Frunzio L,
  Girvin S~M, Devoret M~H and Schoelkopf R~J 2010 {\em Nature\/} {\bf 467}
  574--578 \urlprefix\url{https://doi.org/10.1038/nature09416}

\bibitem{sun2010tunable}
Sun G, Wen X, Mao B, Chen J, Yu Y, Wu P and Han S 2010 {\em Nature
  Communications\/} {\bf 1} 51

\bibitem{Mariantoni2011Implementing}
Mariantoni M, Wang H, Yamamoto T, Neeley M, Bialczak R~C, Chen Y, Lenander M,
  Lucero E, O{\textquoteright}Connell A~D, Sank D, Weides M, Wenner J, Yin Y,
  Zhao J, Korotkov A~N, Cleland A~N and Martinis J~M 2011 {\em Science\/} {\bf
  334} 61--65 ISSN 0036-8075
  \urlprefix\url{https://science.sciencemag.org/content/334/6052/61}

\bibitem{ReedNature2012}
Reed M~D, DiCarlo L, Nigg S~E, Sun L, Frunzio L, Girvin S~M and Schoelkopf R~J
  2012 {\em Nature\/} {\bf 482} 382--385
  \urlprefix\url{https://doi.org/10.1038/nature10786}

\bibitem{cao2013ultrafast}
Cao G, Li H~O, Tu T, Wang L, Zhou C, Xiao M, Guo G~C, Jiang H~W and Guo G~P
  2013 {\em Nature Communications\/} {\bf 4} 1401

\bibitem{thiele2014electrically}
Thiele S, Balestro F, Ballou R, Klyatskaya S, Ruben M and Wernsdorfer W 2014
  {\em Science\/} {\bf 344} 1135--1138

\bibitem{martinis2014fast}
Martinis J~M and Geller M~R 2014 {\em Physical Review A\/} {\bf 90} 022307

\bibitem{wang2018landau}
Wang Z, Huang W~C, Liang Q~F and Hu X 2018 {\em Scientific reports\/} {\bf 8}
  7920

\bibitem{rol2019fast}
Rol M, Battistel F, Malinowski F, Bultink C, Tarasinski B, Vollmer R, Haider N,
  Muthusubramanian N, Bruno A, Terhal B {\em et~al.\/} 2019 {\em Physical
  Review Letters\/} {\bf 123} 120502

\bibitem{sillanpaa2006continuous}
Sillanp{\"a}{\"a} M, Lehtinen T, Paila A, Makhlin Y and Hakonen P 2006 {\em
  Physical Review Letters\/} {\bf 96} 187002

\bibitem{oliver2005mach}
Oliver W~D, Yu Y, Lee J~C, Berggren K~K, Levitov L~S and Orlando T~P 2005 {\em
  Science\/} {\bf 310} 1653--1657

\bibitem{dupont2013coherent}
Dupont-Ferrier E, Roche B, Voisin B, Jehl X, Wacquez R, Vinet M, Sanquer M and
  De~Franceschi S 2013 {\em Physical Review Letters\/} {\bf 110} 136802

\bibitem{nalbach2013nonequilibrium}
Nalbach P, Kn{\"o}rzer J and Ludwig S 2013 {\em Physical Review B\/} {\bf 87}
  165425

\bibitem{huang2011landau}
Huang P, Zhou J, Fang F, Kong X, Xu X, Ju C and Du J 2011 {\em Physical Review
  X\/} {\bf 1} 011003

\bibitem{hicke2006fault}
Hicke C, Santos L and Dykman M 2006 {\em Physical Review A\/} {\bf 73} 012342

\bibitem{bason2012high}
Bason M~G, Viteau M, Malossi N, Huillery P, Arimondo E, Ciampini D, Fazio R,
  Giovannetti V, Mannella R and Morsch O 2012 {\em Nature Physics\/} {\bf 8}
  147--152

\bibitem{gasparinetti2011geometric}
Gasparinetti S, Solinas P and Pekola J~P 2011 {\em Physical Review Letters\/}
  {\bf 107} 207002

\bibitem{tan2014demonstration}
Tan X, Zhang D~W, Zhang Z, Yu Y, Han S and Zhu S~L 2014 {\em Physical Review
  Letters\/} {\bf 112} 027001

\bibitem{zhang2014realization}
Zhang J, Zhang J, Zhang X and Kim K 2014 {\em Physical Review A\/} {\bf 89}
  013608

\bibitem{wang2016experimental}
Wang L, Tu T, Gong B, Zhou C and Guo G~C 2016 {\em Scientific reports\/} {\bf
  6} 19048

\bibitem{zhuang2021noiseresistant}
Zhuang F, Zeng J, Economou S~E and Barnes E 2021 Noise-resistant landau-zener
  sweeps from geometrical curves (\textit{Preprint} \eprint{2103.07586})

\bibitem{Levien:EECS-2008-111}
Levien R 2008 The euler spiral: a mathematical history Tech. Rep.
  UCB/EECS-2008-111 EECS Department, University of California, Berkeley
  \urlprefix\url{http://www2.eecs.berkeley.edu/Pubs/TechRpts/2008/EECS-2008-111.html}

\bibitem{Bartholdi2012}
Bartholdi L and Henriques A 2012 {\em The Mathematical Intelligencer\/} {\bf
  34} 1--3 \urlprefix\url{https://doi.org/10.1007/s00283-012-9304-1}

\bibitem{MatteoOSA2013}
Cherchi M, Ylinen S, Harjanne M, Kapulainen M and Aalto T 2013 {\em Opt.
  Express\/} {\bf 21} 17814--17823
  \urlprefix\url{http://www.opticsexpress.org/abstract.cfm?URI=oe-21-15-17814}

\bibitem{LiLSA2018}
Li L, Lin H, Qiao S, Huang Y~Z, Li J~Y, Michon J, Gu T, Alosno-Ramos C, Vivien
  L, Yadav A, Richardson K, Lu N and Hu J 2018 {\em Light: Science \&
  Applications\/} {\bf 7} 17138--17138
  \urlprefix\url{https://doi.org/10.1038/lsa.2017.138}

\bibitem{weiner1977closed}
Weiner J~L 1977 {\em Proceedings of the American mathematical society\/} {\bf
  67} 306--308

\bibitem{Buterakos_PRXQ2021}
Buterakos D, Das~Sarma S and Barnes E 2021 {\em PRX Quantum\/} {\bf 2}(1)
  010341 \urlprefix\url{https://link.aps.org/doi/10.1103/PRXQuantum.2.010341}

\bibitem{Koch_PRA07}
Koch J, Yu T~M, Gambetta J, Houck A~A, Schuster D~I, Majer J, Blais A, Devoret
  M~H, Girvin S~M and Schoelkopf R~J 2007 {\em Phys. Rev. A\/} {\bf 76}(4)
  042319 \urlprefix\url{https://link.aps.org/doi/10.1103/PhysRevA.76.042319}

\bibitem{Economou_PRB15}
Economou S~E and Barnes E 2015 {\em Phys. Rev. B\/} {\bf 91}(16) 161405

\bibitem{Deng_PRB17}
Deng X~H, Barnes E and Economou S~E 2017 {\em Phys. Rev. B\/} {\bf 96}(3)
  035441 \urlprefix\url{https://link.aps.org/doi/10.1103/PhysRevB.96.035441}

\bibitem{McKay_PRL19}
McKay D~C, Sheldon S, Smolin J~A, Chow J~M and Gambetta J~M 2019 {\em Phys.
  Rev. Lett.\/} {\bf 122}(20) 200502
  \urlprefix\url{https://link.aps.org/doi/10.1103/PhysRevLett.122.200502}

\bibitem{Magesan_PRA20}
Magesan E and Gambetta J~M 2020 {\em Phys. Rev. A\/} {\bf 101}(5) 052308
  \urlprefix\url{https://link.aps.org/doi/10.1103/PhysRevA.101.052308}

\bibitem{Collodo_arxiv20}
Collodo M~C, Herrmann J, Lacroix N, Andersen C~K, Remm A, Lazar S, Besse J~C,
  Walter T, Wallraff A and Eichler C 2020 Implementation of conditional-phase
  gates based on tunable zz-interactions (\textit{Preprint}
  \eprint{2005.08863})

\bibitem{Shulman_Science12}
Shulman M~D, Dial O~E, Harvey S~P, Bluhm H, Umansky V and Yacoby A 2012  {\bf
  336} 202

\bibitem{Wardrop_PRB14}
Wardrop M~P and Doherty A~C 2014 {\em Phys. Rev. B\/} {\bf 90}(4) 045418
  \urlprefix\url{https://link.aps.org/doi/10.1103/PhysRevB.90.045418}

\bibitem{Wang_NPJQI15}
Wang X, Barnes E and Sarma S~D 2015 {\em npj Quantum Information\/} {\bf 1}
  15003 \urlprefix\url{https://doi.org/10.1038/npjqi.2015.3}

\bibitem{Nichol_NPJQI17}
Nichol J~M, Orona L~A, Harvey S~P, Fallahi S, Gardner G~C, Manfra M~J and
  Yacoby A 2017 {\em npj Quantum Information\/} {\bf 3} 3
  \urlprefix\url{https://doi.org/10.1038/s41534-016-0003-1}

\bibitem{CalderonVargas_PRB19}
Calderon-Vargas F~A, Barron G~S, Deng X~H, Sigillito A~J, Barnes E and Economou
  S~E 2019 {\em Phys. Rev. B\/} {\bf 100}(3) 035304
  \urlprefix\url{https://link.aps.org/doi/10.1103/PhysRevB.100.035304}

\bibitem{Qiao_arxiv06}
Qiao H, Kandel Y~P, Dyke J~S~V, Fallahi S, Gardner G~C, Manfra M~J, Barnes E
  and Nichol J~M 2020 Floquet-enhanced spin swaps (\textit{Preprint}
  \eprint{2006.10913})

\bibitem{Krantz_APR19}
Krantz P, Kjaergaard M, Yan F, Orlando T~P, Gustavsson S and Oliver W~D 2019
  {\em Applied Physics Reviews\/} {\bf 6} 021318 (\textit{Preprint}
  \eprint{https://doi.org/10.1063/1.5089550})
  \urlprefix\url{https://doi.org/10.1063/1.5089550}

\bibitem{Malinowski_PRL17}
Malinowski F~K, Martins F, Cywi\ifmmode~\acute{n}\else \'{n}\fi{}ski L, Rudner
  M~S, Nissen P~D, Fallahi S, Gardner G~C, Manfra M~J, Marcus C~M and Kuemmeth
  F 2017 {\em Phys. Rev. Lett.\/} {\bf 118}(17) 177702
  \urlprefix\url{https://link.aps.org/doi/10.1103/PhysRevLett.118.177702}

\bibitem{Fowler_PRA12}
Fowler A~G, Mariantoni M, Martinis J~M and Cleland A~N 2012 {\em Phys. Rev.
  A\/} {\bf 86}(3) 032324
  \urlprefix\url{https://link.aps.org/doi/10.1103/PhysRevA.86.032324}

\bibitem{bikun}
Li B, Calderon-Vargas F~A, Zeng J and Barnes E 2021 Designing arbitrary
  single-axis rotations robust against perpendicular time-dependent noise
  (\textit{Preprint} \eprint{2103.08506})

\bibitem{Liu_PRA_Nogo_softcutoff}
Wang Z~Y and Liu R~B 2013 {\em Phys. Rev. A\/} {\bf 87}(4) 042319
  \urlprefix\url{https://link.aps.org/doi/10.1103/PhysRevA.87.042319}

\bibitem{Reed_PRL2016}
Reed M~D, Maune B~M, Andrews R~W, Borselli M~G, Eng K, Jura M~P, Kiselev A~A,
  Ladd T~D, Merkel S~T, Milosavljevic I, Pritchett E~J, Rakher M~T, Ross R~S,
  Schmitz A~E, Smith A, Wright J~A, Gyure M~F and Hunter A~T 2016 {\em Phys.
  Rev. Lett.\/} {\bf 116}(11) 110402
  \urlprefix\url{https://link.aps.org/doi/10.1103/PhysRevLett.116.110402}

\bibitem{Viola_PRL03}
Viola L and Knill E 2003 {\em Phys. Rev. Lett.\/} {\bf 90} 037901

\bibitem{Brown_PRA2004}
Brown K~R, Harrow A~W and Chuang I~L 2004 {\em Phys. Rev. A\/} {\bf 70}(5)
  052318 \urlprefix\url{https://link.aps.org/doi/10.1103/PhysRevA.70.052318}

\bibitem{Khodjasteh_PRL2009}
Khodjasteh K and Viola L 2009 {\em Phys. Rev. Lett.\/} {\bf 102}(8) 080501
  \urlprefix\url{https://link.aps.org/doi/10.1103/PhysRevLett.102.080501}

\bibitem{Merrill_Wiley14}
Merrill J~T and Brown K~R 2014 {\em vol. 154 (ed. S. Kais), John Wiley \& Sons,
  Inc.\/}
  \urlprefix\url{https://onlinelibrary.wiley.com/doi/abs/10.1002/9781118742631.ch10}

\bibitem{Throckmorton2019}
Throckmorton R~E and Das~Sarma S 2019 {\em Phys. Rev. B\/} {\bf 99}(4) 045422
  \urlprefix\url{https://link.aps.org/doi/10.1103/PhysRevB.99.045422}

\bibitem{Barnes_SciRep15}
Barnes E, Wang X and {Das Sarma} S 2015 {\em Sci. Rep.\/} {\bf 5} 12685

\bibitem{Gungordu2019}
G\"ung\"ord\"u U and Kestner J~P 2019 {\em Phys. Rev. A\/} {\bf 100}(6) 062310
  \urlprefix\url{https://link.aps.org/doi/10.1103/PhysRevA.100.062310}

\bibitem{dong2021doubly}
Dong W, Zhuang F, Economou S~E and Barnes E 2021 Doubly geometric quantum
  control (\textit{Preprint} \eprint{2103.08029})

\bibitem{Barnes_PRL12}
Barnes E and {Das Sarma} S 2012 {\em Phys.\ Rev.\ Lett.\/} {\bf 109} 060401
  \urlprefix\url{https://link.aps.org/doi/10.1103/PhysRevLett.109.060401}

\bibitem{Barnes_PRA13}
Barnes E 2013 {\em Phys.\ Rev.\ A\/} {\bf 88} 013818
  \urlprefix\url{https://link.aps.org/doi/10.1103/PhysRevA.88.013818}

\bibitem{ZANARDI199994}
Zanardi and Rasetti 1999 {\em Physics Letters A\/} {\bf 264} 94 -- 99 ISSN
  0375-9601
  \urlprefix\url{http://www.sciencedirect.com/science/article/pii/S0375960199008038}

\bibitem{Berry.Geometric.Phase}
Berry M~V 1984 {\em Proceedings of the Royal Society of London. A. Mathematical
  and Physical Sciences\/} {\bf 392} 45--57
  \urlprefix\url{https://doi.org/10.1098/rspa.1984.0023}

\bibitem{ANANDAN1988171}
Anandan J 1988 {\em Physics Letters A\/} {\bf 133} 171 -- 175 ISSN 0375-9601
  \urlprefix\url{http://www.sciencedirect.com/science/article/pii/0375960188910109}

\bibitem{Wilczek.Zee}
Wilczek F and Zee A 1984 {\em Phys. Rev. Lett.\/} {\bf 52}(24) 2111--2114
  \urlprefix\url{https://link.aps.org/doi/10.1103/PhysRevLett.52.2111}

\bibitem{Berry_2009}
Berry M~V 2009 {\em Journal of Physics A: Mathematical and Theoretical\/} {\bf
  42} 365303
  \urlprefix\url{https://iopscience.iop.org/article/10.1088/1751-8113/42/36/365303}

\bibitem{SolinasPRA2004}
Solinas P, Zanardi P and Zangh\`{\i} N 2004 {\em Phys. Rev. A\/} {\bf 70}(4)
  042316 \urlprefix\url{https://link.aps.org/doi/10.1103/PhysRevA.70.042316}

\bibitem{Sjoqvist.IJQC.2015}
Sjoqvist E 2015 {\em International Journal of Quantum Chemistry\/} {\bf 115}
  1311--1326
  \urlprefix\url{https://onlinelibrary.wiley.com/doi/abs/10.1002/qua.24941}

\bibitem{Xu.PRL.2012}
Xu G~F, Zhang J, Tong D~M, Sj\"oqvist E and Kwek L~C 2012 {\em Phys. Rev.
  Lett.\/} {\bf 109}(17) 170501
  \urlprefix\url{https://link.aps.org/doi/10.1103/PhysRevLett.109.170501}

\bibitem{Utkan.JPSJ.2014}
Gungordu U, Wan Y and Nakahara M 2014 {\em Journal of the Physical Society
  of Japan\/} {\bf 83} 034001
  \urlprefix\url{https://doi.org/10.7566/JPSJ.83.034001}

\bibitem{PhysRevLett.58.1593}
Aharonov Y and Anandan J 1987 {\em Phys. Rev. Lett.\/} {\bf 58}(16) 1593--1596
  \urlprefix\url{https://link.aps.org/doi/10.1103/PhysRevLett.58.1593}

\bibitem{Sj_qvist_2012}
Sjoqvist E, Tong D~M, Andersson L~M, Hessmo B, Johansson M and Singh K 2012
  {\em New Journal of Physics\/} {\bf 14} 103035
  \urlprefix\url{https://iopscience.iop.org/article/10.1088/1367-2630/14/10/103035}

\bibitem{Sjoqvist}
Sjoqvist E 2016 {\em Physics Letters A\/} {\bf 380} 65 -- 67 ISSN 0375-9601
  \urlprefix\url{http://www.sciencedirect.com/science/article/pii/S0375960115008610}

\bibitem{XueZhengyuan.PRA.2018}
Hong Z~P, Liu B~J, Cai J~Q, Zhang X~D, Hu Y, Wang Z~D and Xue Z~Y 2018 {\em
  Phys. Rev. A\/} {\bf 97}(2) 022332
  \urlprefix\url{https://link.aps.org/doi/10.1103/PhysRevA.97.022332}

\bibitem{Ribeiro.PRA.2019}
Ribeiro H and Clerk A~A 2019 {\em Phys. Rev. A\/} {\bf 100}(3) 032323
  \urlprefix\url{https://link.aps.org/doi/10.1103/PhysRevA.100.032323}

\bibitem{LiuBoajie.PRL.2019}
Liu B~J, Song X~K, Xue Z~Y, Wang X and Yung M~H 2019 {\em Phys. Rev. Lett.\/}
  {\bf 123}(10) 100501
  \urlprefix\url{https://link.aps.org/doi/10.1103/PhysRevLett.123.100501}

\bibitem{YingZuJian.PRR.2020}
Ying Z~J, Gentile P, Baltan\'as J~P, Frustaglia D, Ortix C and Cuoco M 2020
  {\em Phys. Rev. Research\/} {\bf 2}(2) 023167
  \urlprefix\url{https://link.aps.org/doi/10.1103/PhysRevResearch.2.023167}

\bibitem{Shkolnikov.PRB.2020}
Shkolnikov V~O, Mauch R and Burkard G 2020 {\em Phys. Rev. B\/} {\bf 101}(15)
  155306 \urlprefix\url{https://link.aps.org/doi/10.1103/PhysRevB.101.155306}

\bibitem{li2020dynamically}
Li S and Xue Z~Y 2020 Dynamically corrected nonadiabatic holonomic quantum
  gates (\textit{Preprint} \eprint{2012.09034})

\bibitem{Ji2021noncyclic}
Ji L~N, Ding C~Y, Chen T and Xue Z~Y 2021 Noncyclic and nonadiabatic geometric
  quantum gates with smooth paths (\textit{Preprint} \eprint{2102.00893})

\bibitem{Zhao.holonomicDD.PRL.2021}
Zhao P~Z, Wu X and Tong D~M 2021 {\em Phys. Rev. A\/} {\bf 103}(1) 012205
  \urlprefix\url{https://link.aps.org/doi/10.1103/PhysRevA.103.012205}

\bibitem{YanTongxing.PRL.2019}
Yan T, Liu B~J, Xu K, Song C, Liu S, Zhang Z, Deng H, Yan Z, Rong H, Huang K,
  Yung M~H, Chen Y and Yu D 2019 {\em Phys. Rev. Lett.\/} {\bf 122}(8) 080501
  \urlprefix\url{https://link.aps.org/doi/10.1103/PhysRevLett.122.080501}

\bibitem{SunLuYan.PRL.2020}
Xu Y, Hua Z, Chen T, Pan X, Li X, Han J, Cai W, Ma Y, Wang H, Song Y~P, Xue Z~Y
  and Sun L 2020 {\em Phys. Rev. Lett.\/} {\bf 124}(23) 230503
  \urlprefix\url{https://link.aps.org/doi/10.1103/PhysRevLett.124.230503}

\bibitem{Duan.Science.2001}
Duan L~M, Cirac J~I and Zoller P 2001 {\em Science\/} {\bf 292} 1695--1697 ISSN
  0036-8075
  \urlprefix\url{https://science.sciencemag.org/content/292/5522/1695}

\bibitem{AiMingzhong.PRApp.2020}
Ai M~Z, Li S, Hou Z, He R, Qian Z~H, Xue Z~Y, Cui J~M, Huang Y~F, Li C~F and
  Guo G~C 2020 {\em Phys. Rev. Applied\/} {\bf 14}(5) 054062
  \urlprefix\url{https://link.aps.org/doi/10.1103/PhysRevApplied.14.054062}

\bibitem{PhysRevLett.119.140503}
Zhou B~B, Jerger P~C, Shkolnikov V~O, Heremans F~J, Burkard G and Awschalom D~D
  2017 {\em Phys. Rev. Lett.\/} {\bf 119}(14) 140503
  \urlprefix\url{https://link.aps.org/doi/10.1103/PhysRevLett.119.140503}

\bibitem{Zhou2016}
Zhou B~B, Baksic A, Ribeiro H, Yale C, Heremans J, Jerger P, Auer A, Burkard G,
  Clerk A~A and Awschalom D~D 2016 {\em Nature Physics\/} {\bf 13} 330 EP --
  \urlprefix\url{https://doi.org/10.1038/nphys3967}

\bibitem{Sekiguchi.2017}
Sekiguchi Y, Niikura N, Kuroiwa R, Kano H and Kosaka H 2017 {\em Nature
  Photonics\/} {\bf 11} 309--314 ISSN 1749-4893
  \urlprefix\url{https://doi.org/10.1038/nphoton.2017.40}

\bibitem{Mousolou.PRA.2014}
Mousolou V~A and Sj\"oqvist E 2014 {\em Phys. Rev. A\/} {\bf 89}(2) 022117
  \urlprefix\url{https://link.aps.org/doi/10.1103/PhysRevA.89.022117}

\bibitem{PhysRevB.74.205415}
Economou S~E, Sham L~J, Wu Y and Steel D~G 2006 {\em Phys. Rev. B\/} {\bf
  74}(20) 205415
  \urlprefix\url{https://link.aps.org/doi/10.1103/PhysRevB.74.205415}

\bibitem{EconomouPRL2007}
Economou S~E and Reinecke T~L 2007 {\em Phys. Rev. Lett.\/} {\bf 99}(21) 217401
  \urlprefix\url{https://link.aps.org/doi/10.1103/PhysRevLett.99.217401}

\bibitem{EconomouPRB2012}
Economou S~E 2012 {\em Phys. Rev. B\/} {\bf 85}(24) 241401
  \urlprefix\url{https://link.aps.org/doi/10.1103/PhysRevB.85.241401}

\bibitem{Greilich2009}
Greilich A, Economou S~E, Spatzek S, Yakovlev D~R, Reuter D, Wieck A~D,
  Reinecke T~L and Bayer M 2009 {\em Nature Physics\/} {\bf 5} 262--266
  \urlprefix\url{https://doi.org/10.1038/nphys1226}

\bibitem{KuPRA2017}
Ku H~S, Long J~L, Wu X, Bal M, Lake R~E, Barnes E, Economou S~E and Pappas D~P
  2017 {\em Phys. Rev. A\/} {\bf 96}(4) 042339
  \urlprefix\url{https://link.aps.org/doi/10.1103/PhysRevA.96.042339}

\bibitem{Yang_NatElec19}
Yang C~H, Chan K~W, Harper R, Huang W, Evans T, Hwang J~C~C, Hensen B, Laucht
  A, Tanttu T, Hudson F~E, Flammia S~T, Itoh K~M, Morello A, Bartlett S~D and
  Dzurak A~S 2019 {\em Nature Electronics\/} {\bf 2} 151--158
  \urlprefix\url{https://doi.org/10.1038/s41928-019-0234-1}

\bibitem{konnov1999global}
Konnov A and Krotov V~F 1999 {\em Automation and Remote Control\/} {\bf 60}
  1427--1436

\bibitem{palao2002quantum}
Palao J~P and Kosloff R 2002 {\em Phys. Rev. Lett.\/} {\bf 89} 188301

\bibitem{Khaneja_2005}
Khaneja N, Reiss T, Kehlet C, Schulte-Herbruggen T and Glaser S~J 2005 {\em J.
  Magn. Reson.\/} {\bf 172} 296 -- 305

\bibitem{brif2010control}
Brif C, Chakrabarti R and Rabitz H 2010 {\em New J. Phys.\/} {\bf 12} 075008

\bibitem{doria2011optimal}
Doria P, Calarco T and Montangero S 2011 {\em Phys. Rev. Lett.\/} {\bf 106}
  190501

\bibitem{caneva2011chopped}
Caneva T, Calarco T and Montangero S 2011 {\em Phys. Rev. A\/} {\bf 84} 022326

\bibitem{glaser2015training}
Glaser S~J, Boscain U, Calarco T, Koch C~P, K{\"o}ckenberger W, Kosloff R,
  Kuprov I, Luy B, Schirmer S, Schulte-Herbr{\"u}ggen T {\em et~al.\/} 2015
  {\em Eur. Phys. J. D\/} {\bf 69} 279

\bibitem{suter2016colloquium}
Suter D and {\'A}lvarez G~A 2016 {\em Rev. Mod. Phys.\/} {\bf 88} 041001

\bibitem{van2017robust}
Van~Damme L, Ansel Q, Glaser S and Sugny D 2017 {\em Phys. Rev. A\/} {\bf 95}
  063403

\bibitem{Tian_PRA2020}
Tian J, Liu H, Liu Y, Yang P, Betzholz R, Said R~S, Jelezko F and Cai J 2020
  {\em Phys. Rev. A\/} {\bf 102}(4) 043707
  \urlprefix\url{https://link.aps.org/doi/10.1103/PhysRevA.102.043707}

\end{thebibliography}

\end{document}